\documentclass[%
reprint,
superscriptaddress,
twocolumn,
amsmath,amssymb,
aps,
floatfix,
showkeys,
reprint,
]{revtex4-1}

\usepackage{graphicx}
\usepackage{dcolumn}
\usepackage{amsmath}
\usepackage{amssymb}
\usepackage{mathtools}
\usepackage{multirow}
\usepackage{color}
\usepackage{float}
\usepackage{booktabs}
\usepackage{tabularx}
\usepackage[english]{babel}
\usepackage[utf8]{inputenc}
\usepackage{enumerate}
\usepackage[cmyk,dvipsnames]{xcolor}
\usepackage{textcomp}
\usepackage[colorlinks = true,
                        citecolor = blue,
                        linkcolor = blue,
                        urlcolor = blue]{hyperref}
\usepackage[normalem]{ulem}
\usepackage{multirow,tabularx}
\newcolumntype{Y}{>{\centering\arraybackslash}X}

\newcommand{\norm}[1]{\left\lVert#1\right\rVert}
\DeclareMathOperator*{\argmin}{arg\,min}

\begin{document}

\title{Lifetime of coexisting sub-10 nm zero-field skyrmions and antiskyrmions}

\author{Moritz A. Goerzen}
\email[Corresponding author: ]{goerzen@physik.uni-kiel.de}
\affiliation{Institute of Theoretical Physics and Astrophysics, University of Kiel, Leibnizstrasse 15, 24098 Kiel, Germany}

\author{Stephan von Malottki}
\affiliation{Institute of Theoretical Physics and Astrophysics, University of Kiel, Leibnizstrasse 15, 24098 Kiel, Germany}
\affiliation{Science Institute, University of Iceland, 107 Reykjav\'ik, Iceland}
\affiliation{Thayer School of Engineering, Dartmouth College, Hanover, New Hampshire, USA}

\author{Sebastian Meyer}
\affiliation{Institute of Theoretical Physics and Astrophysics, University of Kiel, Leibnizstrasse 15, 24098 Kiel, Germany}
\affiliation{Nanomat/Q-mat/CESAM Université de Liège, B-4000 Sart Tilman, Belgium}

\author{Pavel F. Bessarab}
\affiliation{Science Institute, University of Iceland, 107 Reykjav\'ik, Iceland}
\affiliation{Department of Physics and Electrical Engineering, Linnaeus University, SE-39231 Kalmar, Sweden}

\author{Stefan Heinze}
\affiliation{Institute of Theoretical Physics and Astrophysics, University of Kiel, Leibnizstrasse 15, 24098 Kiel, Germany}
\affiliation{Kiel Nano, Surface, and Interface Science (KiNSIS), University of Kiel, Germany}

\date{\today}

\begin{abstract}
Magnetic skyrmions have raised high hopes for future spintronic devices. For many applications it would be of great advantage to have more than one metastable particle-like texture available. The coexistence of skyrmions and antiskyrmions has been proposed in inversion symmetric magnets with exchange frustration. However, so far only model systems have been studied and the lifetime of coexisting metastable topological spin structures has not been obtained. Here, we predict that skyrmions and antiskyrmions with diameters below 10 nm can coexist at zero magnetic field in a Rh/Co bilayer on the Ir(111) surface -- an experimentally feasible system. We show that the lifetimes of metastable skyrmions and antiskyrmions in the ferromagnetic ground state are above one hour for temperatures up to 75 K and 48 K, respectively. The entropic contribution to the nucleation and annihilation rates differs for skyrmions and antiskyrmions. This opens the route to thermally activated creation of coexisting skyrmions and antiskyrmions in frustrated magnets with Dzyaloshinskii-Moriya interaction.
\end{abstract}

\maketitle

The observation of magnetic skyrmions~\cite{Muehlbauer2009,Yu2010,Heinze2011,Romming2013} has triggered the vision to use these magnetic entities in applications \cite{nagaosa2013topological,Fert2017} ranging from data storage \cite{Fert2013} and logic devices to neuromorphic \cite{Song2020, pinna2018skyrmion, grollier2020neuromorphic} or quantum computing \cite{Psaroudaki2021}. The Dzyaloshinskii-Moriya interaction (DMI) which occurs due to spin-orbit coupling in materials with broken inversion symmetry permits the stabilization of skyrmions at transition-metal interfaces such as ultrathin films at surfaces~\cite{Romming2013,Romming2015,Herve2018,Meyer2019} or magnetic multilayers~\cite{MoreauLuchaire2016,Boulle2016,Woo2016}. In these systems, the variation of chemical composition, stacking sequence, and interface structure opens the possibility to tune the magnetic interactions and thereby skyrmion properties. This approach led to the realization of skyrmions with diameters down to 30~nm at room temperature and the demonstration of current-induced skyrmion motion~\cite{MoreauLuchaire2016,Boulle2016,Woo2016,Soumyanarayanan2017}.

A key limitation of using only skyrmions as information carriers, e.g. in racetrack memory devices \cite{Jiang2017, Litzius2017}, is that the classical bit can only be stored by the presence (`1') or absence (`0') of a skyrmion, which could be impractical since precise control over the skyrmion position is required in this case. It is highly desirable to have another kind of localized, metastable spin structure which can act as a bit in a binary logic \cite{Hoffmann2017,Goebel2021}. The use of distinct magnetic solitons such as degenerate skyrmions (Sk) and antiskyrmions (Ask) has been also proposed for quantum computing~\cite{Psaroudaki2021}. Therefore, systems hosting several co-existing metastable spin structures of different type are not only fundamentally interesting, but also relevant for applications.

A zoo of topological spin structures beyond skyrmions has been theoretically predicted
\cite{Leonov2015,Rozsa2017,Hoffmann2017,Goebel2019,Rybakov2019,Kuchkin2020} and some of them have already been observed in experiments (see e.g.~the review by G\"obel {\it et al.} \cite{Goebel2021}). In tetragonal inverse Heusler compounds antiskyrmions with diameters of about 150 nm can be stabilized in an external magnetic field due to anisotropic DMI and were observed at room temperature by Lorentz transmission electron microscopy \cite{Nayak2017}. The long lifetime of antiskyrmions in these bulk materials was reproduced based on transition-state theory \cite{Potkina2020}. In such materials with anisotropic DMI, which can also be realized at interfaces \cite{Hoffmann2017} and in two-dimensional magnets \cite{Ga2022}, antiskyrmions are energetically preferred over skyrmions. By use of in-plane magnetic fields, it was experimentally possible to transform antiskyrmions into elliptical skyrmions in this class of materials \cite{Jena2020,Peng2020,karube2022high}. The observed coexistence phase of two magnetic objects with opposite topological charge $Q$
arises in these systems from the competition of anisotropic Dzyaloshinskii-Moriya and dipole-dipole interactions \cite{Koshibae2016}. In ferrimagnetic multilayers, even purely dipolar-interaction stabilized skyrmions and antiskyrmions with diameters of about 200 nm have been observed \cite{Heigl2021}. It has also been shown that stable skyrmion and antiskyrmion solutions of the continuum model coexist in a limited range of magnetic field and/or anisotropy \cite{Kuchkin2020a} even without magnetostatic interaction. Recently, the creation and annihilation of such skyrmion-antiskyrmion pairs with diameters of about 150~nm have been observed in thin films of the cubic chiral magnet FeGe \cite{Zheng2022}. However, co-existing nanoscale skyrmions and antiskyrmions have not been reported so far.

In order to realize coexisting sub-10 nm skyrmions and antiskyrmions which are ideally suited to outperform competing future technologies \cite{Fert2017} the interplay of other magnetic interactions can be exploited. It has been shown that frustrated exchange \cite{Leonov2015,Lin2016,Malottki2017,Zhang2017,Desplat2019} as well as higher-order exchange \cite{Paul2020} interactions can stabilize isolated Sk and Ask in the field-polarized phase of
a material, i.e.~at finite magnetic field, even without DMI. However, the lifetime and relative probability of coexisting metastable topological spin structures has not been reported so far.

Here, we predict the coexistence of isolated skyrmions and antiskyrmions with diameters below 10 nm at zero magnetic field in an atomic Rh/Co bilayer on the Ir(111) surface -- an experimentally feasible ultrathin film system~\cite{Meyer2019,Perini2019}. Due to strong exchange frustration, both types of topological spin structures are stable down to zero magnetic field, i.e.~in the ferromagnetic ground state of the system. We calculate the lifetimes of metastable skyrmions and antiskyrmions using transition-state theory based on an atomistic spin model with all interactions parameterized from density functional theory (DFT). Lifetimes above one hour are obtained for skyrmions and antiskyrmions in Rh/Co/Ir(111) at temperatures of up to 75~K and 48~K, respectively. The lifetimes of antiskyrmions are even longer than those of skyrmions in the prototypical skyrmion system Pd/Fe/Ir(111) \cite{Romming2013,dupe2014tailoring,Hagemeister2015,Malottki2019,Muckel2021}.
 
Further we study the effect of DMI on skyrmions and antiskyrmions in a frustrated magnet, in particular, on the energy barriers and pre-exponential factors in the Arrhenius law describing kinetics of skyrmion and antiskyrmion annihilation and nucleation. We demonstrate that the difference in the energy spectra for skyrmions and antiskyrmions in presence of DMI leads to distinct entropic contributions to their transition rates. Surprisingly, even in the regime of significant DMI the creation of coexisting skyrmions and antiskyrmions becomes possible via local heating.

\section*{}
\noindent{\large{\textbf{Results}}}\par
\noindent{\textbf{Zero-field skyrmions and antiskyrmions.}} 
We study skyrmions and antiskyrmions in an atomic Rh/Co bilayer on the Ir(111) surface using the corresponding
DFT-parameterized atomistic spin model \cite{Meyer2019} given by
\begin{eqnarray}
    E = &-&
    \sum_{i,j} J_{ij} (\mathbf{m}_i\cdot\mathbf{m}_j)-
    \sum_{i,j} \mathbf{D}_{ij}\cdot(\mathbf{m}_i\times\mathbf{m}_j) \nonumber \\
    &-&\sum_{i}K(\mathbf{m}_i\cdot\hat{\mathbf{e}}_{\perp})^2-\sum_{i}
    M (\mathbf{m}_i\cdot\mathbf{{\cal B}}),
    \label{eq:spin_model}
\end{eqnarray}
which takes into account the Heisenberg exchange interaction and the DMI between normalized magnetic moments $\mathbf{m}_i$ and $\mathbf{m}_j$ at lattice sites $i$ and $j$ of the Co layer as well as the uniaxial magnetocrystalline anisotropy and the Zeemann energy due to an external magnetic field $\mathbf{{\cal B}}$ applied perpendicular to the film. The exchange constants, $J_{ij}$, the strength of the DMI, $D_{ij}$, the magnetocrystalline anisotropy constant, $K$, and the size of the magnetic moment per site, $M$, have been obtained based on DFT for both fcc- and hcp-stacking of the Rh layer \cite{Meyer2019}(see methods for details and Supplementary Note 1 and Supplementary Figure 1). 

\begin{figure}
    \centering
    \includegraphics[width=0.49\textwidth, keepaspectratio]{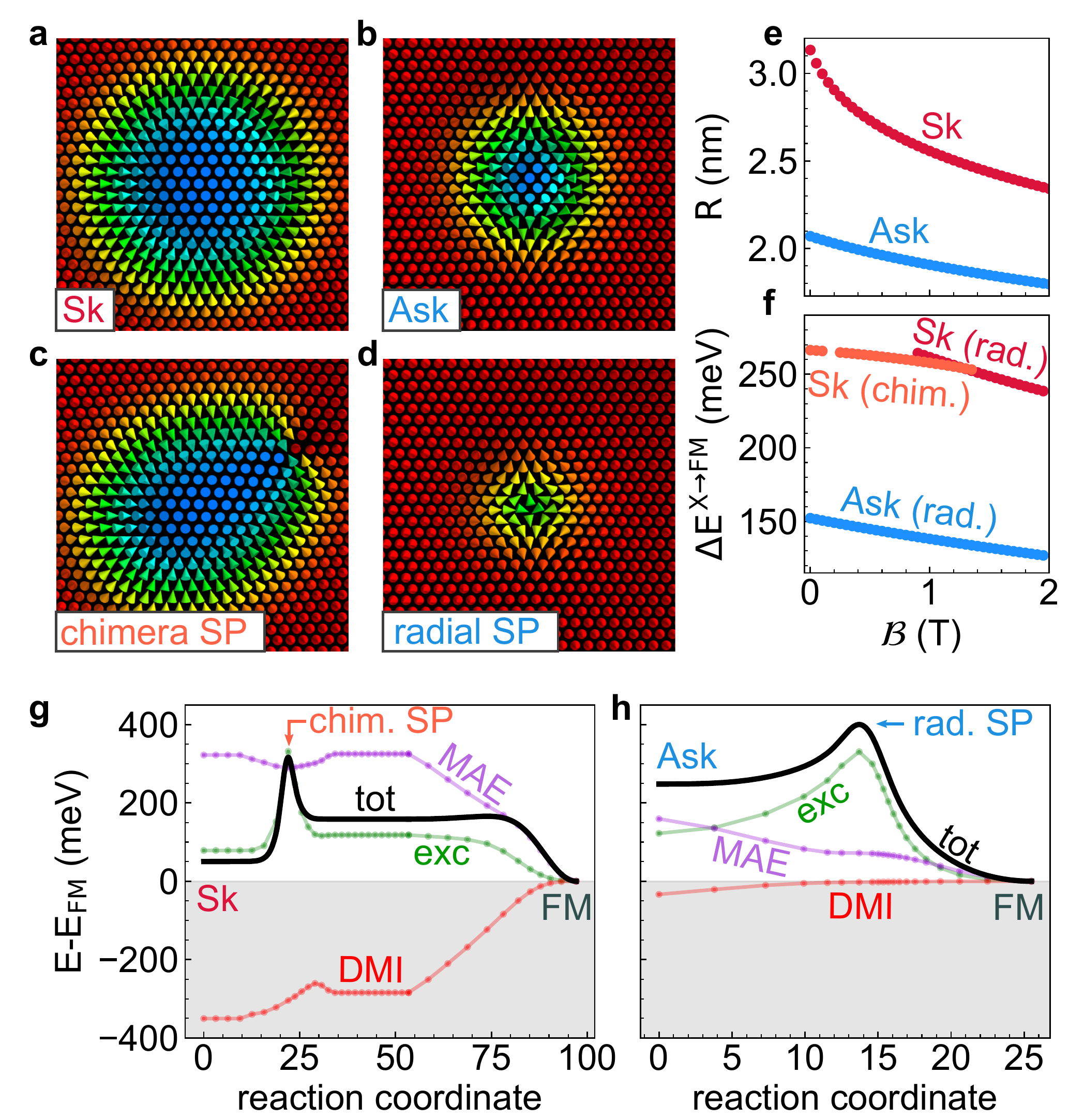}
    \caption{\textbf{Zero-field skyrmions and antiskyrmions in fcc-Rh/Co/Ir(111).} \textbf{a} Isolated skyrmion (Sk) and \textbf{b} isolated antiskyrmion (Ask) at ${\cal B}=0$~T. \textbf{c} Chimera (skyrmion) saddle point (SP) and \textbf{d} antiskyrmion SP. Images \textbf{a-d} show equally sized parts of the much larger respective simulation boxes. \textbf{e} Radius of skyrmions (Sk, red) and antiskyrmions (Ask, blue) as a function of magnetic field applied perpendicular to the film. \textbf{f} Energy barriers $\Delta E$ for skyrmion (Sk$\to$FM, red) and antiskyrmion (Ask$\to$FM, blue) annihilation. For skyrmions, there is a crossover from the Chimera collapse (orange) at low fields to the radial collapse (red) mechanism at larger magnetic fields. Minimum energy path (MEP) for \textbf{g} the collapse of a skyrmion (initial state) via the Chimera mechanism into the FM (final) state and \textbf{h} the antiskyrmion collapse via the radially-symmetric mechanism at ${\cal B}=0$ T. The total energy (tot), with respect to the FM ground state, is decomposed into the contributions from the exchange interaction (exc), the DMI, and the magnetocrystalline anisotropy energy (MAE).}
    \label{fig:zero_field}
\end{figure}

Isolated skyrmions and antiskyrmions (Fig.~\ref{fig:zero_field}a,b) are initialised within the ferromagnetic ground state (for simulation details, such as initial skyrmion profiles and optimizations, see methods). The radii of the skyrmions and antiskyrmions are obtained from the relaxed profiles using the definition of Bogdanov \textit{et al.}~\cite{bocdanov1994properties}. Even in the absence of magnetic field both types of spin structures are stable in our simulations and exhibit diameters well below 10~nm (Fig.~\ref{fig:zero_field}e). Note that nanoscale skyrmions of this size have already been experimentally observed in Rh/Co/Ir(111) at zero magnetic field \cite{Meyer2019,Perini2019}.

Antiskyrmions are by about 20 to 50\% smaller than skyrmions in the entire range of magnetic fields due to the effect of DMI. In order to study the stability of both states, we calculate the energy barriers protecting them against collapse into the ferromagnetic ground state using the geodesic nudged elastic band (GNEB) method \cite{bessarab2015method} (see methods). As presented in Fig.~\ref{fig:zero_field}f we find that the skyrmion annihilation energy barrier $\Delta E^{\text{Sk}\to\text{FM}}$ is significantly larger than that for antiskyrmions, $\Delta E^{\text{Ask}\to\text{FM}}$, for all magnetic fields. Surprisingly, the antiskyrmion barrier of about 150~meV at ${\cal B}=0$~T is of the same order of magnitude as the barrier for the skyrmion collapse reported for Pd/Fe/Ir(111) \cite{Malottki2017}, a well-studied system for which experiment~\cite{Romming2013,Romming2015,Hagemeister2015,Lindner2020,Muckel2021} and first-principles based 
theory~\cite{dupe2014tailoring,Malottki2017,Boettcher2018,Muckel2021} are in good agreement. This suggests that antiskyrmions in Rh/Co/Ir(111) could possess a similar stability as that observed for skyrmions in Pd/Fe/Ir(111) at low temperatures~\cite{Romming2013,Romming2015,Hagemeister2015,Lindner2020,Muckel2021}. 

The GNEB calculations also provide information about mechanisms of magnetic transitions. For skyrmions, we find the recently discovered Chimera collapse mechanism \cite{Meyer2019,Muckel2021} at ${\cal B} < 1.26$~T denoted as Sk (chim.) in Fig.~\ref{fig:zero_field}f. At larger fields, the radially symmetric transition, denoted as Sk (rad.), is energetically favoured over the Chimera transition. From the minimum energy path for the Chimera collapse at ${\cal B} =0$~T (Fig.~\ref{fig:zero_field}g), we see that the energy of the skyrmion state (Fig.~\ref{fig:zero_field}a) itself is lowered by the DMI while the energy costs of the saddle point (Fig.~\ref{fig:zero_field}c), which determines the energy barrier, stems mostly from the frustrated exchange interaction.
    
The situation is different for antiskyrmions (Fig.~\ref{fig:zero_field}b). There, we find only the radial shrinking annihilation mechanism in the entire investigated range of magnetic fields in which the initial state (Fig.~\ref{fig:zero_field}b) collapses into the ferromagnetic ground state via a saddle point (Fig.~\ref{fig:zero_field}d) exhibiting a core with in-plane pointing spins. The minimum energy path at ${\cal B}=0$ T (Fig.~\ref{fig:zero_field}h) shows that the energy barrier is dominated by the frustrated exchange interaction. In contrast to skyrmions, the influence of the DMI on the antiskyrmion collapse is very small.
    
The stability of metastable topological spin structures can be quantified by their mean lifetime, $\tau$. Here we use transition state theory in harmonic approximation to the energy of states \cite{bessarab2018lifetime} where the expression for the lifetime takes the form of an Arrhenius law
\begin{equation}\label{eq:Arrhenius}
    \tau=\tau_{0} \exp{(\Delta E/k_{\rm B}T)}
\end{equation}
where $\Delta E$ is the energy barrier of the transition and $\tau_0$ a pre-exponential factor which can be explicitly calculated (see methods).

\begin{figure}
    \centering
    \includegraphics[width=0.355\textwidth, keepaspectratio]{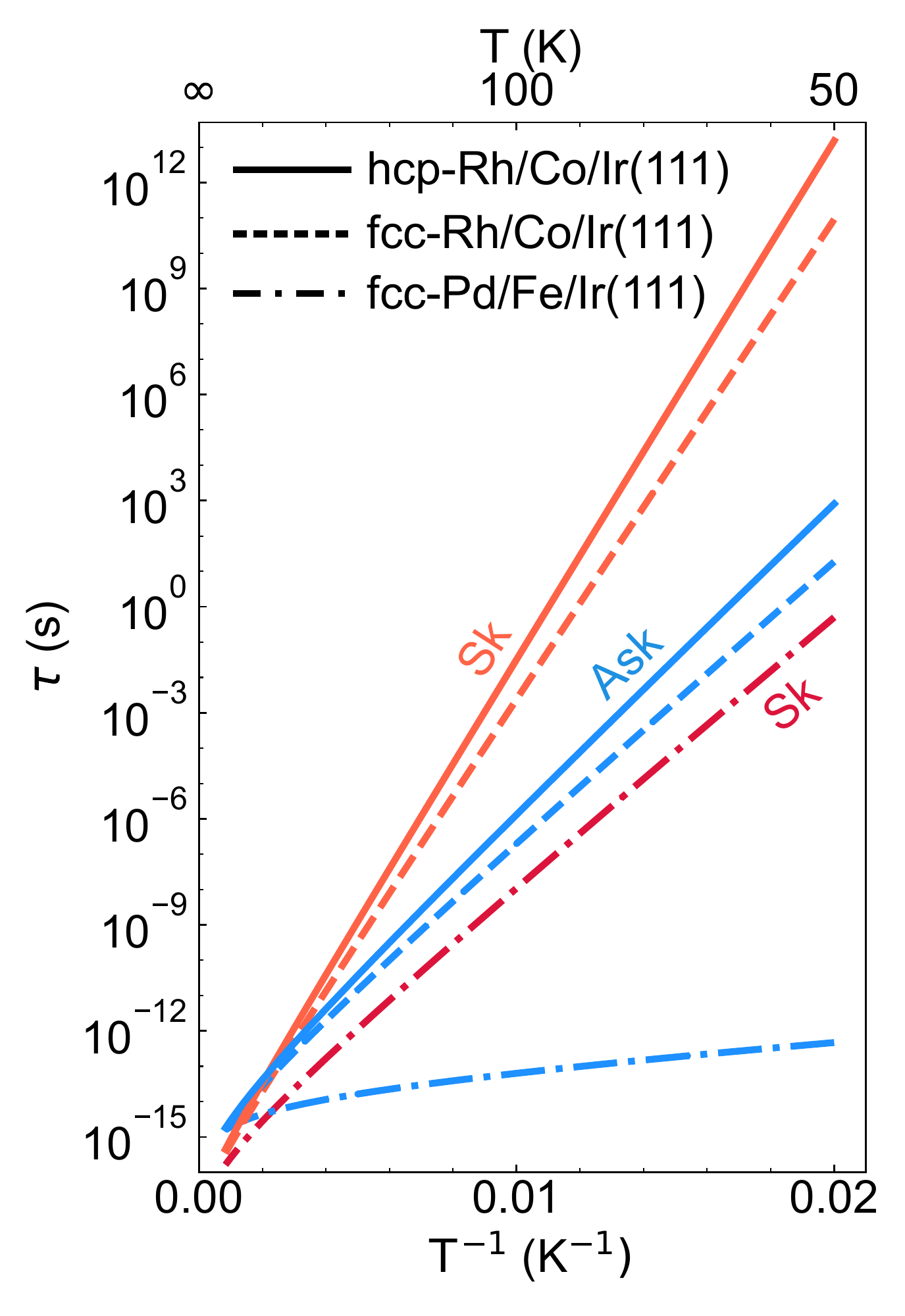}
    \caption{\textbf{Lifetime of skyrmions and antiskyrmions in Rh/Co/Ir(111)}. Lifetime of skyrmions (Sk, orange lines) and antiskyrmions (Ask, blue lines) in hcp-Rh/Co/Ir(111) (solid lines) and fcc-Rh/Co/Ir(111) (dashed lines) at $\mathcal{B}=0~T$ shown over inverse temperature. For comparison the lifetimes are also given for skyrmions and antiskyrmions in Pd/Fe/Ir(111) (dashed-dotted lines) at $\mathcal{B}=3.9~T$, i.e.~just above the critical field of the field-polarized phase (see methods for details).}
    \label{fig:lifetime_cmp}
\end{figure}

The obtained lifetimes of skyrmions and antiskyrmions in Rh/Co/Ir(111) at zero magnetic field are displayed in Fig.~\ref{fig:lifetime_cmp} for both fcc- and hcp-stacking of the Rh layer in a logarithmic plot as functions of inverse temperature. Considering the logarithmic scaling of the ordinate
\begin{equation}\label{eq:Arrhenius_log}
    \ln\tau = \ln\tau_0 + \Delta E/k_BT,
\end{equation}
the slope of the curves thus gives the energy barrier, while the offset at $T^{-1}\to 0$ yields the pre-exponential factor. As expected from the energy barriers, the lifetime of skyrmions is much larger than that of antiskyrmions at low temperatures. Nevertheless, the lifetime of antiskyrmions is also above one hour below about 48~K and 43~K in hcp-Rh/Co and fcc-Rh/Co on Ir(111), respectively. For skyrmions, the one-hour lifetime is already exceeded below about 75~K and 66~K for the hcp and fcc stacking of Rh, respectively. Note, that at large temperatures there is a crossing of skyrmion and antiskyrmion lifetime due to different pre-exponential factors.

For comparison, we have also displayed in Fig.~\ref{fig:lifetime_cmp} the calculated lifetimes for isolated metastable skyrmions in the field-polarized phase of Pd/Fe/Ir(111)~\cite{Malottki2019} at $\mathcal{B}=3.9~$T. This value has been chosen because at this field the annihilation barriers for Sk in Pd/Fe/Ir(111) and Ask in Rh/Co/Ir(111) are similar, despite the fact, that the Ask are significantly smaller (see Supplementary Note 2 and Supplementary Figure 2). Remarkably, the lifetime of Ask in Rh/Co/Ir(111) is even larger than that of Sk in Pd/Fe/Ir(111). Note, that the calculated lifetimes \cite{Malottki2019,goerzen2022atomistic} are in good 
agreement with experiments for Pd/Fe/Ir(111) \cite{Hagemeister2015,Muckel2021}. We have also calculated the lifetime of antiskyrmions in Pd/Fe/Ir(111) at $\mathcal{B}=3.9~$T. However, even at the lowest considered temperatures their lifetime is very short (Fig.~\ref{fig:lifetime_cmp}). This can be explained based on the much smaller energy barrier as apparent from the small slope of the curve. The origin of this difference between the two film systems is, in addition to the applied magnetic field, the significantly enhanced exchange frustration in Rh/Co/Ir(111) \cite{Meyer2019} (see Supplementary Information).

\noindent{\textbf{DMI vs.~frustrated exchange.}}
To understand the impact of the DMI on skyrmions and antiskyrmions in a two-dimensional system and its competition with frustrated exchange interactions, we vary the DMI strength in our atomistic spin simulations. The exchange energy is identical for skyrmions and antiskyrmions with the same radial profiles (see methods). In contrast, DMI provides different contributions to the energy of skyrmions and antiskyrmions. For skyrmions, it leads to a preference of N\'eel-type skyrmions with a unique rotational sense due to the symmetry at the interface. For undistorted antiskyrmions the DMI energy vanishes. The origin of this difference is that each radial cut through the skyrmion (Fig.~\ref{fig:zero_field}a) exhibits a clockwise spin rotation favored by the DMI. The antiskyrmion (Fig.~\ref{fig:zero_field}b) exhibits opposite rotational senses along its two major axes which leads to zero total DMI energy for an undistorted antiskyrmion. In our simulations, however, we observe a small elongation of the antiskyrmion along the favored axis, leading to a lowering in energy. The maximal distortion can be seen in Fig.~\ref{fig:zero_field}b, where the favored axis points in the y-direction. Similar but larger elongations of Ask have also been observed in Ref.~\cite{Kuchkin2020a}.
    
In an inversion-symmetric material, the DMI cancels out leading to the degeneracy of skyrmions and antiskyrmions. Model systems of frustrated magnets in the absence of DMI have been studied in the past \cite{Leonov2015,Lin2016}. In the considered Rh/Co bilayer on Ir(111), there is a significant DMI due to the heavy metal substrate which energetically prefers skyrmions over antiskyrmions while the exchange frustration stabilises both. Because of this exchange frustration we found comparatively large lifetimes for both types of topological spin structures (cf. Fig.~\ref{fig:lifetime_cmp}) but the skyrmion lifetimes exceed those of antiskyrmions due the DMI.

Since the DMI strength may vary depending on the considered interface, it is interesting to study the impact of DMI on the transition rates and lifetimes systematically. We parameterise the energy term due to DMI in our 
atomistic spin model, Eq.~(\ref{eq:spin_model}), by $\eta \in [0,1]$:
\begin{equation}
    E_{\rm DMI} (\eta) = -\eta\sum_{i,j}\mathbf{D}_{ij}\cdot(\mathbf{m}_i\times\mathbf{m}_j)
\end{equation}
such that $\eta=1$ describes the full atomistic spin model for fcc-Rh/Co/Ir(111) as obtained from DFT and $\eta=0$ an inversion symmetric film with frustrated exchange but without DMI (for similar results obtained for hcp-Rh/Co/Ir(111) see Supplementary Note 3 and Supplementary Figure 3).

\begin{figure}
    \centering
    \includegraphics[width=0.5\textwidth, keepaspectratio]{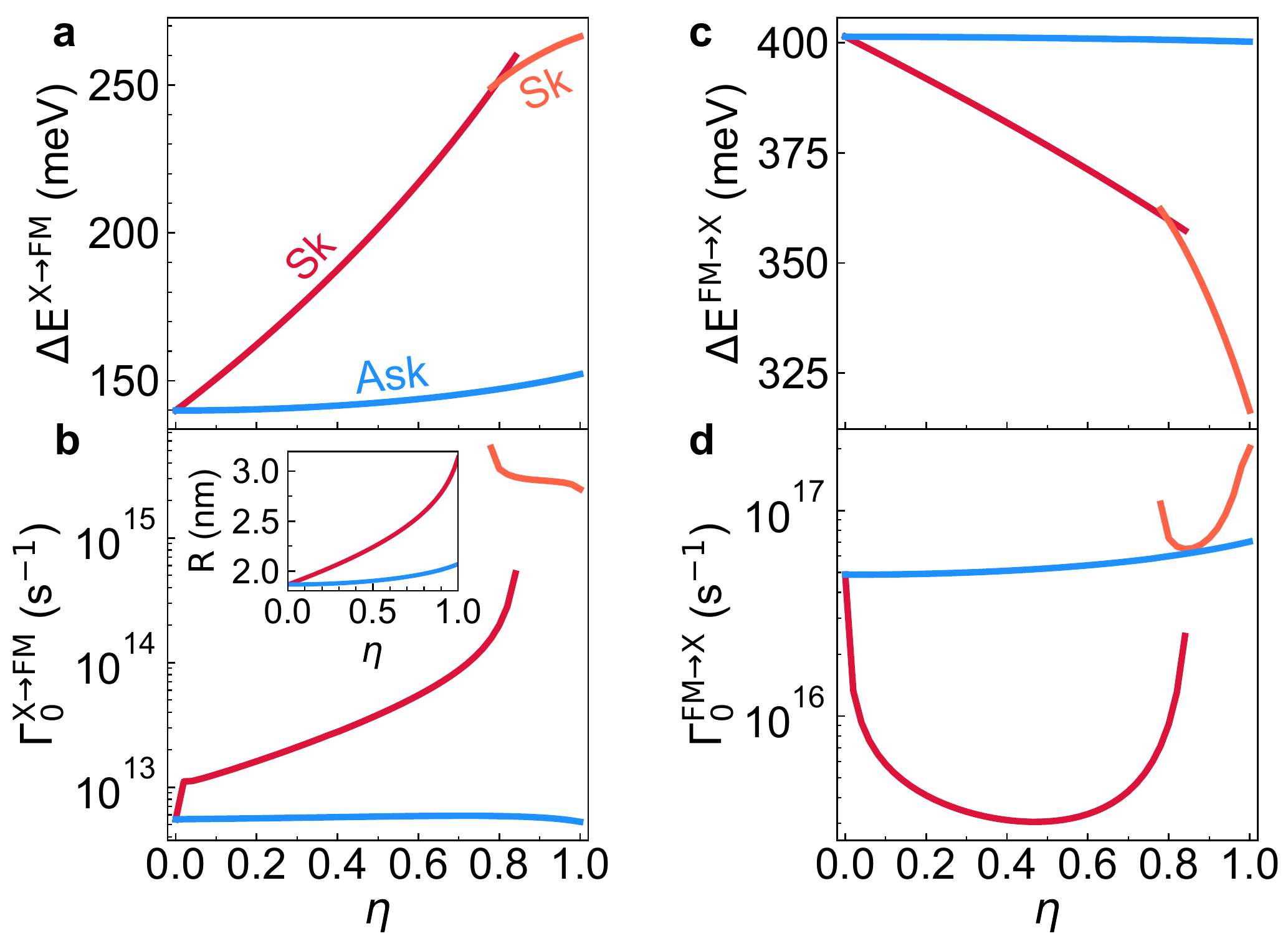}
    \caption{\textbf{Energy barriers and pre-exponential factors vs.~DMI strength.} \textbf{a} Energy barriers for skyrmion (Sk, red and orange) and antiskyrmion (Ask, blue) annihilation (X$\to$FM with X$\in\{\text{Sk},\text{Ask}\}$) as a function of DMI strength $\eta$. \textbf{b} Pre-exponential factors for skyrmion and antiskyrmion annihilation at $T=1~$K. \textbf{c} Activation energies for skyrmion and antiskyrmion nucleation (FM$\to$X with X$\in\{\text{Sk},\text{Ask}\}$) and \textbf{d} corresponding pre-exponential factors at $T=1~$K. Here the helicity partition functions in the pre-exponential factor for skyrmion annihilation and nucleation shown in \textbf{b},\textbf{d} have been calculated numerically using the temperature $T=1~$K (see methods). This is essential in order to assure continuity between harmonic and zero mode approximation at $\eta\to0$. In all panels, the red curves correspond to the radially symmetric skyrmion transition mechanism while the orange curves correspond to the Chimera mechanism. The inset in \textbf{b} shows the radius of skyrmions (red) and antiskyrmions (blue) as a function of $\eta$.}
    \label{fig:radius}
\end{figure}

So far, we have considered the stability of skyrmions and antiskyrmions versus collapse into the ferromagnetic ground state, i.e.~their mean lifetime $\tau$ as given by the Arrhenius law Eq.~(\ref{eq:Arrhenius}). In order to discuss the annihilation and nucleation of skyrmions or antiskyrmions on equal footing, it is convenient to resort to the rate for a transition between state $A$ and $B$ defined by
\begin{equation} \label{eq:transition_rate}
    \Gamma^{\text{A}\to\text{B}} = \Gamma^{\text{A}\to\text{B}}_0\exp(-\Delta E^{\text{A}\to\text{B}}/k_{\rm B}T),
\end{equation}
where the activation energy $\Delta E^{\text{A}\to\text{B}} = E^{\text{SP}} -E^{\text{A}}$ is obtained as the energy difference between the initial state A and the transition-state SP, and the pre-exponential factor $\Gamma_0^{\text{A}\to\text{B}}$ takes into account the entropic and dynamic contributions to the transition \cite{desplat2018thermal,varentcova2020toward,Malottki2019,goerzen2022atomistic}. Therefore, in our notation $\Gamma^{\text{X}\to\text{FM}}$ denotes the annihilation rate and $\Gamma^{\text{FM}\to\text{X}}$ the nucleation rate for a state X$\in\{\text{Sk},\text{Ask}\}$. Note that the superposition of all transition rates for collapses of a skyrmion or antiskyrmion is the inverse of the respective lifetime $\tau$.
    
From our simulations we find that skyrmions and antiskyrmions are metastable in the full range of $\eta$ (Fig.~\ref{fig:radius}). As expected, skyrmions and antiskyrmions exhibit the same radius (inset of Fig.~\ref{fig:radius}b) for vanishing DMI ($\eta=0$). With increasing DMI, the skyrmion diameter rises, while there is only a very small change for antiskyrmions.

Similar to the radii we observe a degeneracy of the annihilation barriers for skyrmions and antiskyrmions (Fig.~\ref{fig:radius}a) at zero DMI ($\eta= 0$) and a similar collapse via the radial transition mechanism. For increased strength of the DMI ($\eta>0$) the skyrmion collapse barrier rises. At $\eta \approx 0.78$ the Chimera transition mechanism sets in and the slope of the rise changes. For $\eta>0.92$ the Chimera state becomes metastable, which can be seen from the very shallow minimum in the MEP after the saddle point (Fig.~\ref{fig:zero_field}g). However, it exhibits only a tiny energy barrier of $\Delta E< 8$~meV against a collapse into the FM state barely visible on the energy scale of Fig.~\ref{fig:zero_field}g. In relation to all other transitions this energy barrier is negligibly small and therefore it is ignored in the skyrmion annihilation/nucleation rate calculations.

There is only a minor influence of the DMI on the antiskyrmion annihilation energy barrier, as expected from the simplified micromagnetic model (see methods). The small increase in the annihilation barrier with the DMI strength (Fig.~\ref{fig:radius}a) is due to the antiskyrmions' elliptical deformation that decreases their energy.

The DMI-dependent changes in activation energies for the nucleation of skyrmions (FM$\to$Sk) and antiskyrmions (FM$\to$ASk) (Fig.~\ref{fig:radius}c), can be explained in a similar way. For skyrmions we find a lowering of the nucleation barrier with rising DMI, while the barrier is nearly constant for antiskyrmions. This indicates that for low temperatures the nucleation of skyrmions is favored over the nucleation of antiskyrmions in the presence of DMI. 

The obtained pre-exponential factors $\Gamma^{\text{A}\to\text{B}}_0$, essential for the computation of transition rates (Eq.~(\ref{eq:transition_rate})), are shown in Fig.~\ref{fig:radius}b,d for the annihilation and nucleation, respectively. For the annihilation of skyrmions, the pre-exponential factor rises monotonically with $\eta$ for the radial collapse mechanism and changes by about two orders of magnitude. A rising pre-exponential factor increases the likelihood of the respective transition. Therefore, skyrmion annihilation occurs more often with rising $\eta$, effectively decreasing the state's average lifetime. Thus DMI causes entropic destabilisation of skyrmions, opposite to the energetic stabilisation. In the regime of the Chimera collapse, there is a slight drop with the DMI strength. For skyrmion nucleation (Fig.~\ref{fig:radius}d), the pre-exponential factor displays a local minimum at $\eta \approx 0.5$.

In contrast, for antiskyrmions the pre-exponential factors change only slightly with varying DMI strength (Fig.~\ref{fig:radius}b,d). To understand this behavior, we analyze the skyrmion and antiskyrmion eigenmodes.

\begin{figure}
    \centering
    \includegraphics[width=0.49\textwidth, keepaspectratio]{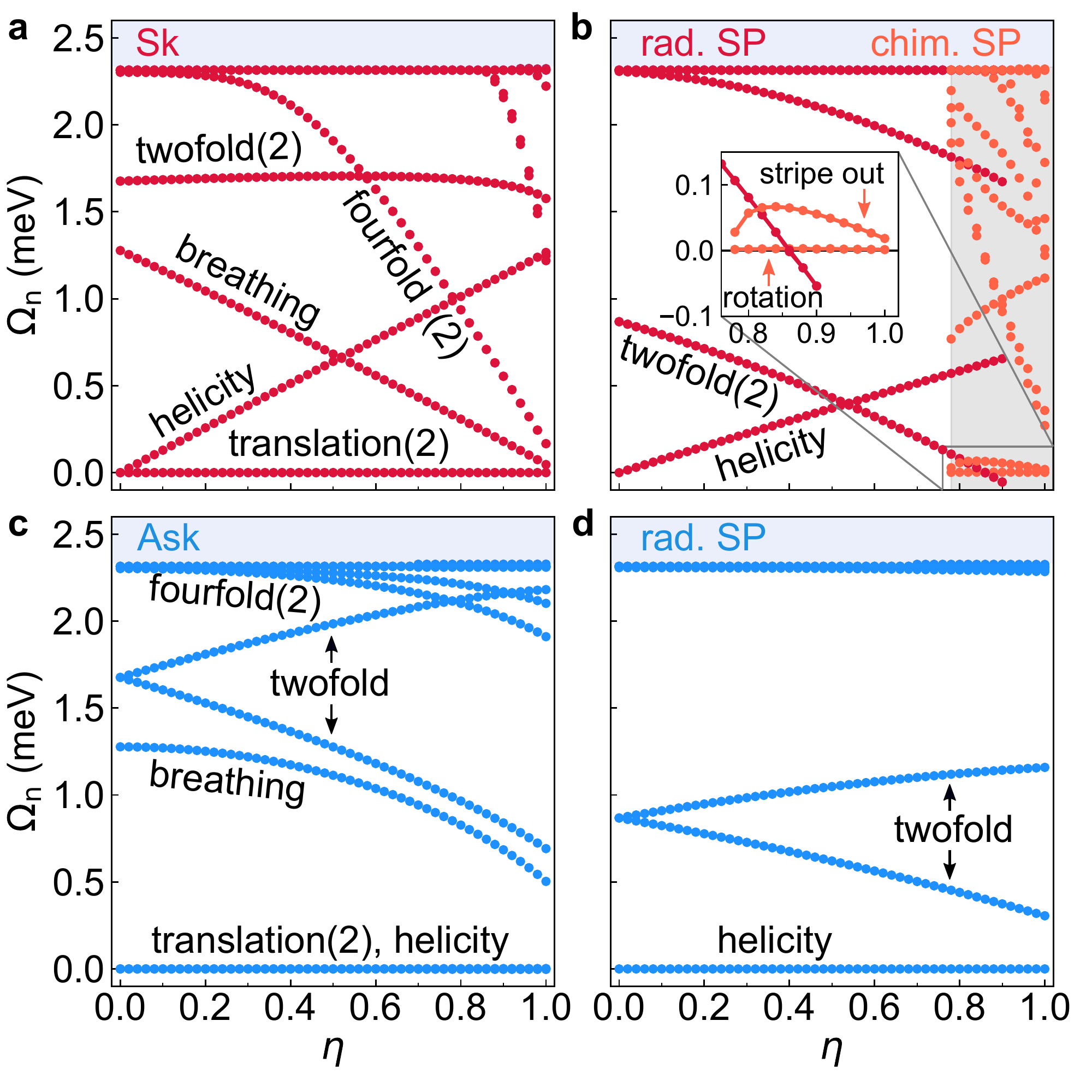}
    \caption{\textbf{Energy modes of skyrmions and antiskyrmions vs.~DMI
    strength.} Eigenvalue spectra as functions of $\eta$: \textbf{a} for skyrmion initial state \textbf{b} for skyrmion saddle point state, radial (red) and chimera (orange) \textbf{c} for antiskyrmion initial state and \textbf{d} antiskyrmion radial saddle point state. Numbers in brackets give the amount of degenerated modes. The shaded blue area on top of each panel marks the magnon continuum with the lower boundary at $\Omega_{\text{mag}}= 2K \approx 2.33$ meV.}
    \label{fig:eigenvalues}
\end{figure}

\noindent{\textbf{Skyrmion and antiskyrmion modes.}} 
Within harmonic transition state theory (HTST), the entropic contribution to the pre-exponential factor is given by the ratio of partition functions for the initial state and the transition state defined by the saddle point \cite{bessarab2018lifetime,varentcova2020toward}, where the partition functions are inversely proportional to the square root of the eigenvalues of the Hessian at the given state (see methods). Therefore, we analyze the eigenvalue spectra of the skyrmion and its saddle point (Fig.~\ref{fig:eigenvalues}a,b) as well as the antiskyrmion and its saddle point (Fig.~\ref{fig:eigenvalues}c,d). We focus on eigenvalues, $\Omega_n$, describing the curvature of the energy surface along localized modes below the magnon continuum \cite{potkina2020skyrmions}, i.e.~$0\ \text{meV}\leq \Omega_n \leq 2K \approx 2.33\ \text{meV}$, where $K$ is the MAE constant. We identify the associated modes by analysing the eigenvectors of the Hessian matrix as in Refs.~\cite{Malottki2019,varentcova2020toward, schrautzer2022effects}.

\begin{figure*}[!htbp]
    \centering
    \includegraphics[width=1\textwidth, keepaspectratio]{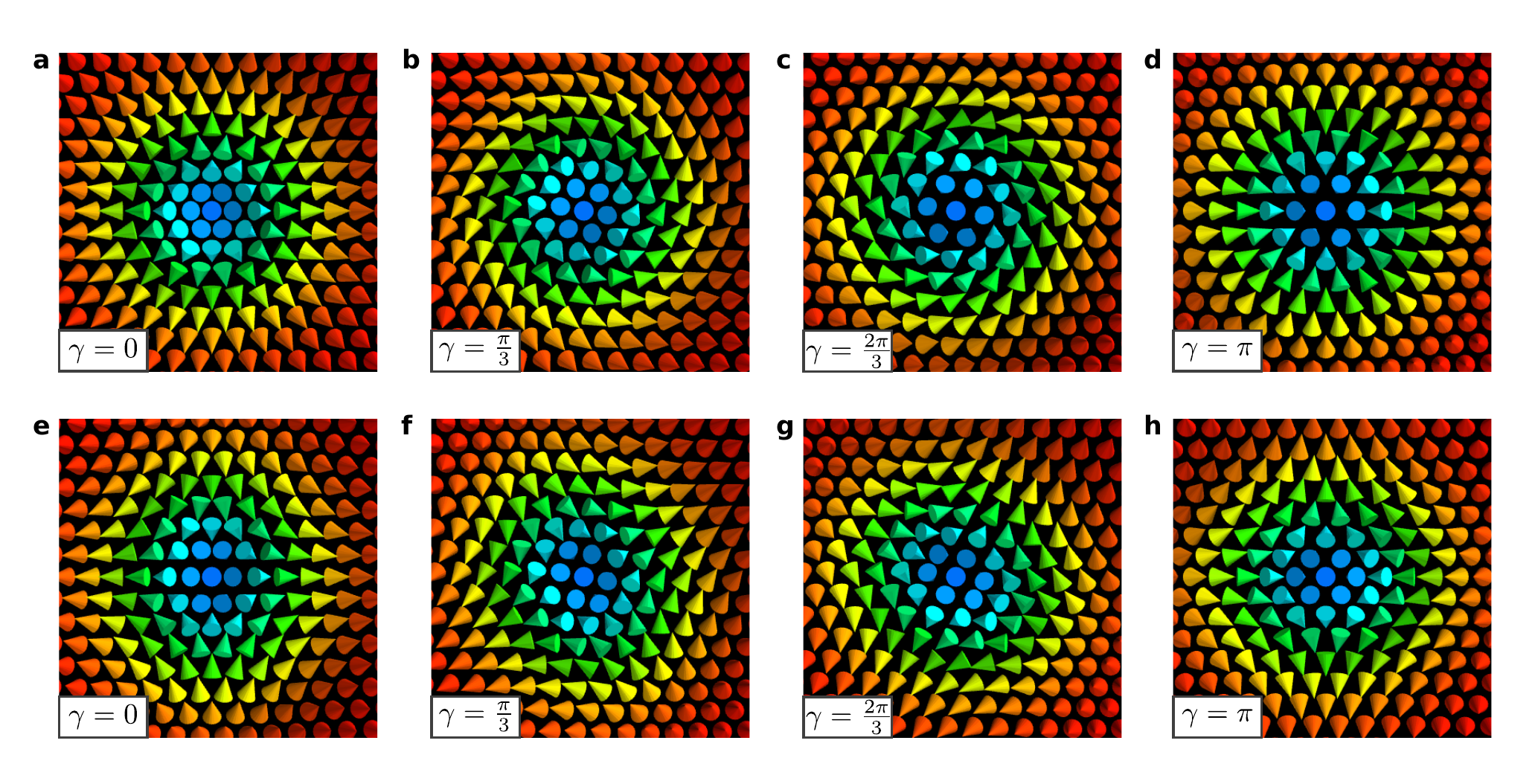}
    \caption{\textbf{Illustration of helicity modes for skyrmions and antiskyrmions.} \textbf{a-d} Transformation of a N\'eel-type skyrmion with a counter-clockwise rotational sense (\textbf{a}, $\gamma=0$) via mixed Bloch-N\'eel skyrmions (\textbf{b}, $\gamma=\frac{\pi}{3}$ and \textbf{c}, $\gamma=\frac{2\pi}{3}$) into a N\'eel-type skyrmion with the opposite rotational sense (\textbf{d}, $\gamma=\pi$). \textbf{e-h} The same sequence of helicities starting from a N\'eel-type antiskyrmion (\textbf{e}) leads to a $90^\circ$ rotation of the antiskyrmion in real space.}
    \label{fig:helicity_concept}
\end{figure*}

As expected, the energy spectra for skyrmions and antiskyrmions coincide for $\eta=0$, i.e.~in the absence of DMI. The same holds for the corresponding saddle point states. With increasing $\eta$, key differences arise:
    
{\it Twofold mode.} The twofold mode \cite{desplat2018thermal,Malottki2019} describes a unidirectional distortion of the state. At the saddle point state, it can be seen as the movement of the Bloch point like defect along a lattice direction. For the skyrmion, as a radial symmetric state, both distortions along different lattice directions increase the energy of the state by the same amount. Therefore, the twofold eigenmodes are degenerate (Fig.~\ref{fig:eigenvalues}a,b). For the antiskyrmion, in contrast, the distortions along the directions, where spins rotate clockwise or counter-clockwise, are only degenerate if the DMI is zero. With rising DMI, the distortion along one direction is favored over the other, which consequently leads to a splitting of the eigenvalues as observed in Fig.~\ref{fig:eigenvalues}c,d.

{\it Fourfold mode.} This mode \cite{Malottki2019, desplat2018thermal} quenches the state along a specific direction and elongates it perpendicular to this direction. As for the twofold mode we find two fourfold modes with interchanged directions for the quenching/elongation, which are degenerate for both skyrmions and antiskyrmions. 

{\it Helicity mode.} The helicity mode \cite{Lin2016,desplat2020entropy} transforms between the N\'eel-type and the Bloch-type configuration of a skyrmion as illustrated in Fig.~\ref{fig:helicity_concept}a-d. Due to DMI the helicity $\gamma$ only affects the energy of skyrmions and not of undistorted antiskyrmions (see methods). In our film system, the DMI favors the N\'eel-type configuration with a right-rotating sense which corresponds to $\gamma = \pi$ (Fig.~\ref{fig:helicity_concept}d). For $\eta=0$, i.e.~in the absence of DMI, N\'eel- and Bloch-type skyrmions are degenerate. Therefore, the mode which transforms one into the other is a zero (Goldstone) mode, i.e.~its eigenvalue is zero (see Fig.~\ref{fig:eigenvalues}a,b). For finite values of DMI strength, the eigenvalue corresponding to the helicity mode becomes positive for skyrmions. This is contrary to antiskyrmions, where the helicity mode stays a zero mode for all values of $\eta$ (Fig.~\ref{fig:eigenvalues}c,d).

The different behavior of skyrmions and antiskyrmions can be understood by looking at their symmetry. Changing the skyrmions helicity results in a twist of the spin structure (Fig.~\ref{fig:helicity_concept}a-d), which costs energy in the presence of DMI. For the antiskyrmion, on the other hand, a change in helicity by $\Delta\gamma$ results in a real space rotation of the spin structure by $\Delta\gamma/2$ (Fig.~\ref{fig:helicity_concept}e-h), leaving the energy of the state unchanged.
    
To our knowledge, the treatment of the helicity zero mode in HTST or Langer's theory has not been reported in the literature so far. For antiskyrmions at arbitrary $\eta$ and skyrmions at $\eta=0$ we present an analytic expression for the zero mode partition function (see methods). For $\eta>0$ we find temperature dependent deviations in the zero mode- and harmonic approximation of the skyrmion helicity compared with numerically computed results (see Supplementary Note 4 and Supplementary Figure 4). For a correct treatment of this mode we therefore use only numerically computed partition functions for the skyrmions and its radial SP's helicity in the regime $\eta>0$ (see methods). Other zero modes are the translation of skyrmions and antiskyrmions treated as in Ref.~\cite{haldar2018first} and the rotation of the chimera saddle point configuration which is determined by numerical integration along the mode, similar to skyrmions helicity for $\eta>0$. With these methods we are able to compute the pre-exponetial factor of skyrmions and antiskyrmions in the whole range of $\eta$.

The eigenvalue spectra  (Fig.~\ref{fig:eigenvalues}) allow us to understand the dependence of the pre-exponential factors on the DMI strength $\eta$ (Fig.~\ref{fig:radius}b,d). The nearly constant values for pre-exponential factors of annihilation and nucleation of antiskyrmions can hardly be understood by looking at individual modes. The DMI is influencing modes all over the spectrum and consequently affecting the corresponding eigenvalues. Its negligible influence on the pre-exponential factor must therefore be seen as a global phenomenon.

For skyrmions, the diverging behaviour of the annihilation pre-exponential factor with $\eta\to1$ can be explained by the behaviour of the saddle point twofold modes. Considering the dependence $\Gamma^{\text{Sk}\to\text{FM}}_0\propto\Omega_{\text{tf}}^{-2}$ with $\Omega_{\text{tf}}$ being the eigenvalue of one of the two degenerated twofold modes, the pre-exponential factor diverges for $\Omega_{\text{tf}}\to0$, which is the case for high $\eta$. The divergence of the skyrmions nucleation pre-exponential factor (Fig.~\ref{fig:radius}d) can be explained similarly.

\noindent{\textbf{Transition rates}.} 
Based on the calculated activation energies and pre-exponential factors (Fig.~\ref{fig:radius}), we can compute the annihilation and nucleation rates of skyrmions and antiskyrmions according to Eq.~(\ref{eq:transition_rate}). The temperature dependence of the rates stems not only from the exponential decay over the energy barrier, but also from the pre-exponential factors, which inherit the temperature dependence from the single modes Boltzmann partition functions (see methods).

\begin{figure}
    \centering
    \includegraphics[width=0.49\textwidth, keepaspectratio]{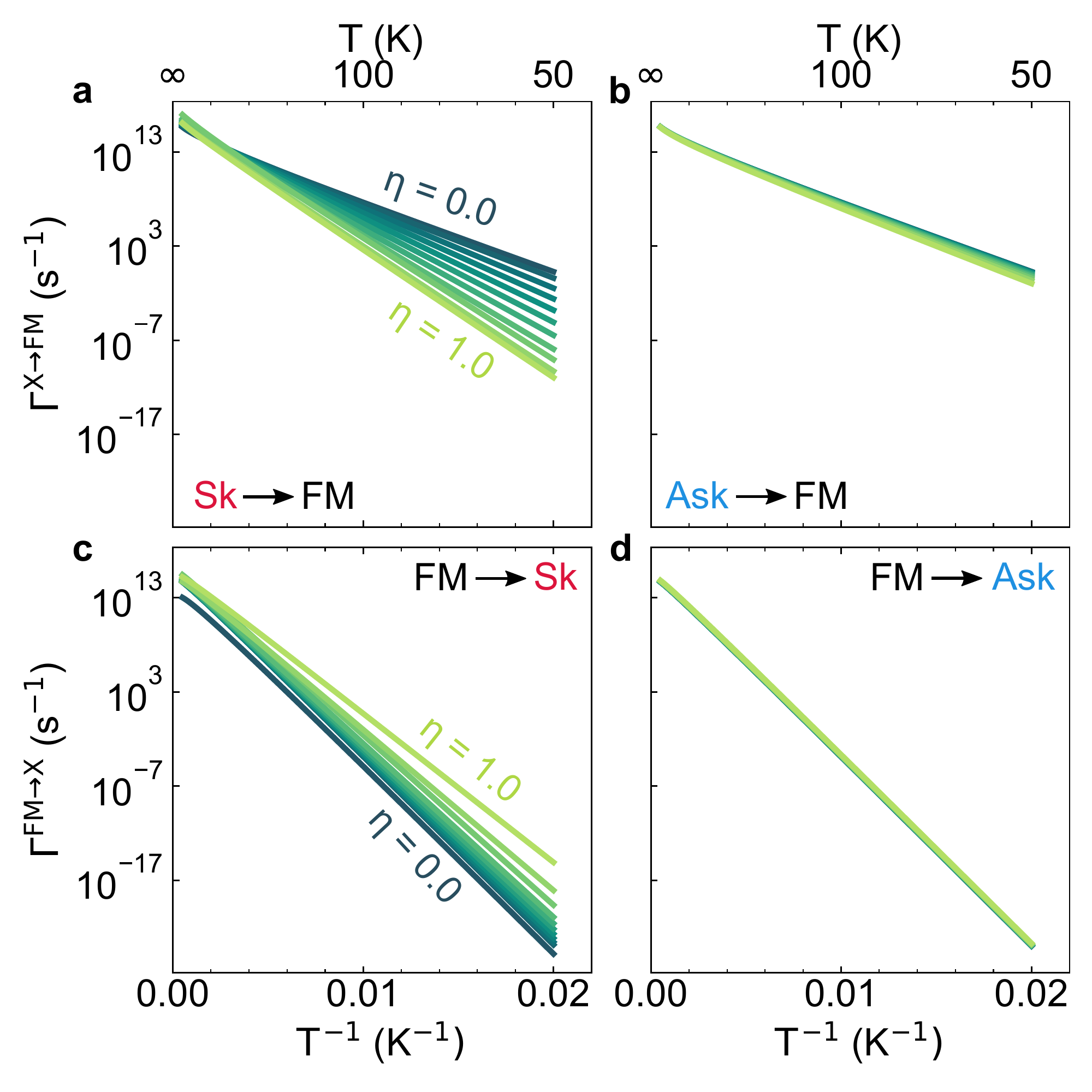}
    \caption{\textbf{Annihilation and nucleation rates of skyrmions and antiskyrmions.} Transition rates of
    \textbf{a} skyrmion and \textbf{b} antiskyrmion annihilation as well as \textbf{c} skyrmion and \textbf{d} antiskyrmion nucleation. All rates are displayed vs.~inverse temperature and for 11 different values of $\eta$ evenly spaced in the interval $[0,1]$. Radial and chimera transition as well as the superposition of both (skyrmion rates for $\eta\in[0.78, 0.84]$) are taken into account at the respective values of $\eta$.}
    \label{fig:lifetimes_rates}
\end{figure}

The annihilation rates (Fig.~\ref{fig:lifetimes_rates}a) range over many orders of magnitude depending on the temperature. We notice that the skyrmion annihilation rates decrease tremendously at low temperatures with rising DMI strength, i.e.~with $\eta$. This effect is due to rising of the annihilation barrier with the DMI strength (Fig.~\ref{fig:radius}a) and is reflected in Fig.~\ref{fig:lifetimes_rates}a by the larger slopes of the annihilation rates for the larger $\eta$ values. The effect of the DMI strength on the antiskyrmion annihilation (Fig.~\ref{fig:lifetimes_rates}b) is quite small in agreement with its impact on the energy barrier (cf.~Fig.~\ref{fig:zero_field}h and Fig.~\ref{fig:radius}a). The slope of the nucleation rate plots (Fig.~\ref{fig:lifetimes_rates}c,d) displays a similar dependence on the DMI strength as the annihilation rates. Considering the logarithmic scaling of the ordinate (cf. Eq.~(\ref{eq:Arrhenius_log})), the linear character of the curves also indicates that the activation energies $\Delta E$ are mostly responsible for the scaling of the rates over the shown range of temperatures. Nevertheless, the pre-exponential factor acts not only as an offset at $T^{-1}\to 0$ but also as small correction to the curves for increasing temperatures due to its intrinsic temperature dependence listed in Tab.~\ref{tab:number_zero_modes}.

Due to the increase of the skyrmions annihilation pre-exponential factor with $\eta$ (cf.~Fig.~\ref{fig:radius}c) this leads to a crossing of annihilation rates for high and low $\eta$ at high temperatures (cf.~Fig.~\ref{fig:lifetimes_rates}a). Considering that the antiskyrmion annihilation rates stay mostly unchanged with $\eta$, this is equal to a crossing of skyrmion and antiskyrmion annihilation rates for high $\eta$. Therefore, we can expect skyrmions to be destroyed more often than antiskyrmions at high temperatures in the realistic system with high DMI. In the next section we point out how this entropic effect influences the equilibrium between skyrmions and antiskyrmions.
    
\begin{figure}
    \centering
    \includegraphics[width=0.48\textwidth, keepaspectratio]{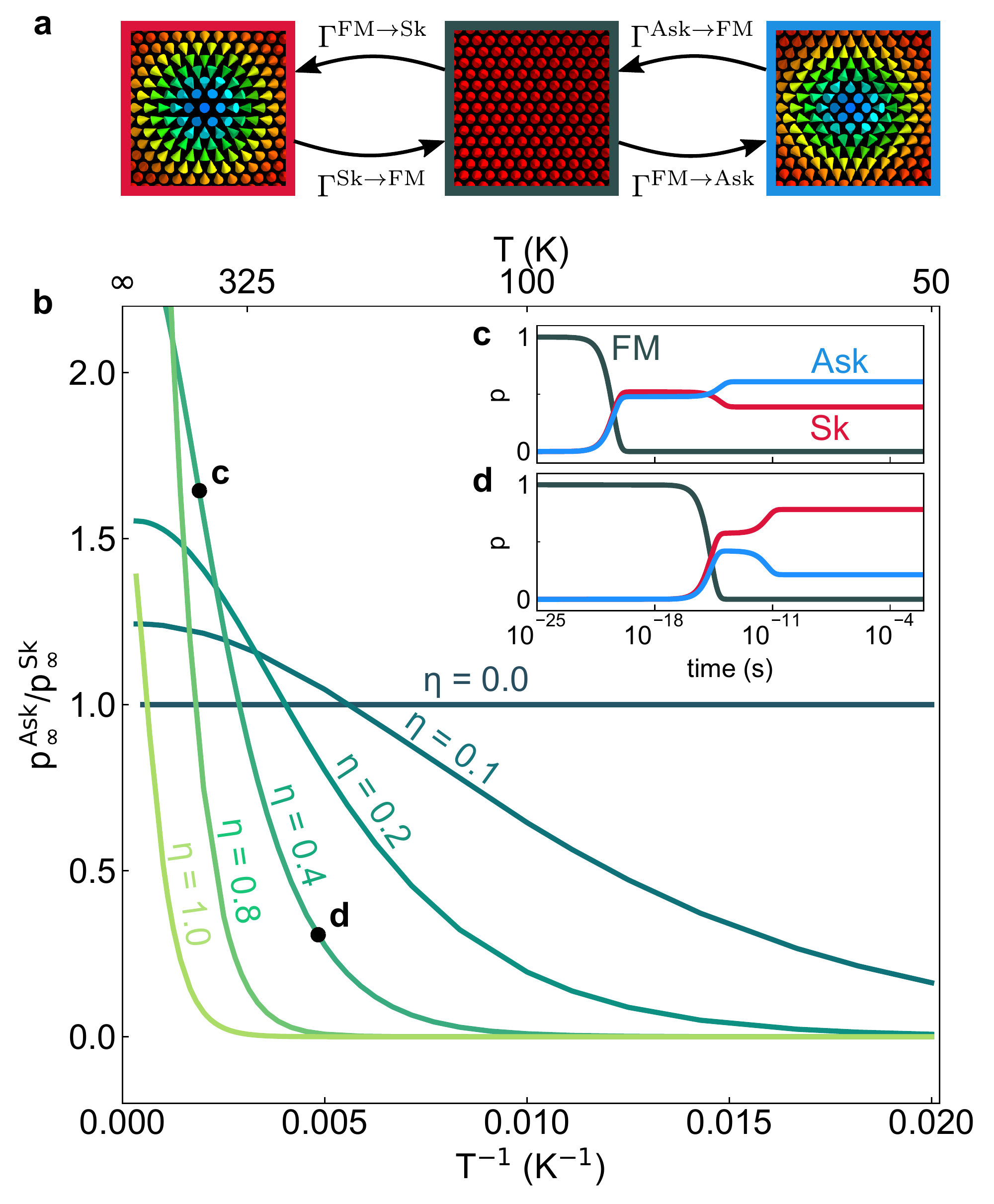}
    \caption{\textbf{Ratio of probability distributions of skyrmions and antiskyrmions in thermal equilibrium.} \textbf{a} Schematic representation of Markov chain containing all transitions considered in the master equations. \textbf{b} The relative probabilities of skyrmions and antiskyrmions in thermal equilibrium are shown for different values of the DMI strength $\eta$. The insets show the time-dependence of the probabilities for the skyrmion, antiskyrmion, and FM state obtained by solving the master equation for $\eta=0.4$ at \textbf{c} $T=500~$K and \textbf{d} $T=200~$K}.
    \label{fig:master_equation}
\end{figure}

\noindent{\textbf{Skyrmion and antiskyrmion probabilities.}} 
We can use the transition rates from Fig.~\ref{fig:lifetimes_rates} to compute the relative probabilities of zero-field skyrmions and antiskyrmions at a given temperature and DMI strength. To do so, we model the problem as a Markov chain as illustrated in Fig.~\ref{fig:master_equation}a and solve a three-state master equation for a system that can be either in the skyrmion, in the antiskyrmion, or in the FM state (see methods for details). Note, that the transitions Sk$\to$ASk and ASk$\to$Sk have not been found in our GNEB calculations of minimum energy paths. These always end in a transition Sk$\to$FM$\to$ASk or vice versa which in turn is included in our description by the master equation.

From the numerical solution of the master equation we obtain the time-dependent probabilities $p^{\rm Sk} (t)$, $p^{\rm ASk} (t)$, and $p^{\rm FM} (t)$ to find the system in the Sk, Ask, or FM state, respectively. Two examples for the time-dependent probabilities are given for a DMI strength of $\eta=0.4$, in Fig.~\ref{fig:master_equation}c,d, one for $T=200~$K, the other for $T=500~$K. Note, that a rule of thumb in HTST, which states the theories validity when $\Delta E \gtrsim 5k_BT$ is fulfilled for all energy barriers, gives us the upper limit $T\lesssim 325~$K considering the lowest energy barrier of $140~$meV for the antiskyrmion annihilation at $\eta=0$. For orientation this limit is denoted as tick in Fig.~\ref{fig:master_equation}b. In both insets Fig.~\ref{fig:master_equation}c,d the FM state is initialised at $t=0$ in order to simulate the nucleation of skyrmions and antiskyrmions. In these examples we find that antiskyrmions and skyrmions are nucleated on a similar time scale. However, for higher temperatures skyrmions decay faster due to the crossing of annihilation rates observed in Fig.~\ref{fig:lifetimes_rates}a. 

By taking the values in the limit $t\to\infty$, we can obtain the equilibrium distribution of the states. These values are independent of the initial state chosen. In Fig.~\ref{fig:master_equation}b, the ratio of the equilibrium probabilities for skyrmions and antiskyrmions is shown. In the absence of DMI ($\eta=0$), skyrmions and antiskyrmions are degenerate and thus the ratio of probabilities is unity. Increasing the DMI strength $\eta$ stabilises skyrmions versus antiskyrmions at low temperatures. However, for high temperatures the antiskyrmion becomes more likely than the skyrmion due to the effect discussed for Fig.~\ref{fig:master_equation}c. This entropic effect becomes even more significant for high values of $\eta$ but the crossover ($p^{\text{Ask}}_{\infty}/p^{\text{Sk}}_{\infty}=1$) also shifts to higher temperatures. Nevertheless it is remarkable that there exists a regime of temperatures below the rule-of-thumb-limit of $T\lesssim325~$K, where antiskyrmions become more likely to be found than skyrmions.

This general effect, which is based on the simultaneous stabilisation of skyrmions and antiskyrmions due to exchange frustration and entropic destabilisation of skyrmions due to the interplay of DMI and temperature, opens the door for thermal manipulation of the relative probability of skyrmions and antiskyrmions. Although their lifetimes drop exponentially with temperatures (cf. Fig.~\ref{fig:lifetimes_rates}), short and local heating pulses, the heat of which dissipates faster than the characteristic time for the equilibration between skyrmions and antiskyrmions (cf.~insets of Fig.\ref{fig:master_equation}b), can nucleate antiskyrmions away from the equilibrium distribution due to the Kibble-Zurek mechanism \cite{zurek1985cosmological, kibble1976topology} which has recently also been proposed based on Monte-Carlo simulations for the creation of skyrmions and antiskyrmions in inversion-symmetric materials \cite{Lin2016}. According to our simulations this effect permits for the thermal nucleation of antiskyrmions even for low global temperatures in frustrated magnets with broken inversion symmetry and significant DMI. This greatly extends the type of material systems in which coexisting metastable skyrmions and antiskyrmions can be created.

\section*{}
\noindent{\large{\textbf{Discussion}}}\par
Based on an atomistic spin model parametrized from DFT we have demonstrated that sub-10 nm skyrmions and antiskyrmions can coexist at zero magnetic field in the ferromagnetic ground state of the Rh/Co bilayer on the Ir(111) surface and that they exhibit large lifetimes of longer than an hour up to temperatures of 50~K. Since skyrmions have already been observed experimentally in Rh/Co/Ir(111)~\cite{Meyer2019}, our prediction may enable the discovery of nanoscale antiskyrmions and their coexistence with skyrmions. Our work further revealed that frustrated magnets even with significant DMI can be a promising class of materials to study the coexistence of spin structures with different topological charge. Such systems can also be realized in two-dimensional van der Waals magnets \cite{Li2022} and in magnetic multilayers \cite{Nickel2023}.

By systematically studying the influence of the DMI strength on the transition rates, we found
that only skyrmions undergo entropic destabilisation with increasing DMI. Due to this effect the relative probability of skyrmions and antiskyrmions can be tuned by thermal manipulation in frustrated magnets with broken inversion symmetry, e.g.~by local heating using lasers or electrical currents. Together with the Kibble-Zurek mechanism this can be exploited for the probabilistic nucleation of antiskyrmions away from equilibrium as needed for neuromorphic or Bayesian computing \cite{pinna2018skyrmion}.

\section*{}\label{sec:methods}
\noindent{\large{\textbf{Methods}}}\par

\noindent{{\textbf{Atomistic spin model of Rh/Co/Ir(111).}}}
For all atomistic spin simulations on Rh/Co/Ir(111) presented in this work, we used the interaction constants given in table \ref{tab:interactions} to parameterize the spin model, Eq.~(\ref{eq:spin_model}). All interaction constants used in our simulations have been obtained previously by means of DFT calculations \cite{Meyer2019}. We have performed simulations using the constants for fcc- and hcp-stacking of the Rh overlayer. Note, that the DMI strength has been reduced to 50 \% of the DFT value ($\eta=0.5$) in case of hcp-Rh/Co/Ir(111) which leads to a very good agreement with experimental data on skyrmions in this system \cite{Meyer2019}. The values of the magnetocrystalline anisotropy energy are $K = - 1.166~$meV and $K=-1.635$~meV for fcc-Rh/Co/Ir(111) and hcp-Rh/Co/Ir(111), respectively, where the negative sign indicates an easy out-of-plane magnetization axis and the magnetic moment is $M=2.5~\mu_{\text{B}}$ per atom. For all calculations we used system sizes of either 50$\times$50 or 70$\times$70 lattice points with periodic boundary conditions representing the magnetic moments of the Co layer.

\begin{table}[]
    \begin{ruledtabular}
    \begin{tabular}{ l c c c c }
        \multicolumn{1}{c}{\multirow{1}{*}{}} & \multicolumn{2}{c}{$J_n$ (meV)} & \multicolumn{2}{c}{$D_n$ (meV)}\\
         & fcc-Rh/Co & hcp-Rh/Co & fcc-Rh/Co & hcp-Rh/Co \\
        \hline
        1 & 25.180 & 29.217 & $-$0.290 & 0.130 \\
        2 & 0.254 & 0.855 & 0.108 & 0.382\\
        3 & $-$2.706 & $-$4.069 & 0.299 & 0.374\\
        4 & $-$0.634 & $-$0.414 & $-$0.030 & $-$0.027\\
        5 & $-$0.237 & $-$0.580 & $-$0.017 & $-$0.239\\
        6 &  0.095 & $-$0.127 & $-$0.038 & $-$0.056\\
        7 &  0.016 & 0.005 & 0.086 & 0.178\\
        8 &  0.273 & 0.047 & $-$ & $-$\\
        9 &  0.030 & 0.048 & $-$ & $-$\\
        10& $-$0.281 & $-$0.159 & $-$ & $-$\\
    \end{tabular}
    \end{ruledtabular}
    \caption{\textbf{DFT interaction constants of Rh/Co/Ir(111)}
    Shell-resolved exchange ($J_n$) and DMI ($D_n$) constants as obtained from DFT calculations, rounded to third digit. Positive values of $D_n$ represent a clockwise rotation preferred by the DMI.}
    \label{tab:interactions}
\end{table}

\noindent{{\textbf{Skyrmion and antiskyrmion profile.}}}
For initialising a skyrmion or antiskyrmion state in our simulation box, we use the following parametrization \cite{nagaosa2013topological}
\begin{equation}\label{eq:skyrmion_ansatz}
    \mathbf{m}(\phi, \rho) = \left( \begin{array}{c}    
    \cos[\Phi(\phi)]\sin[\Theta(\rho)]\\
    \sin[\Phi(\phi)]\sin[\Theta(\rho)]\\
    \cos[\Theta(\rho)]
    \end{array}  \right)
\end{equation}
where $\phi, \rho$ are the polar coordinates with respect to the skyrmion center. The angular functions are given by
\begin{eqnarray}
    \Phi(\phi) &=& k\phi+\gamma  \\
    \Theta(\rho) &=& \pi + \sum_{\alpha=\pm 1} \arcsin\left[\tanh\left(\frac{2(\alpha c + \rho)}{w}\right)\right]
    \label{eq:theta_profile}
\end{eqnarray}
with vorticity $k\in\mathbb{Z}$, helicity $\gamma\in[0,2\pi]$ and profile parameters $c,w>0$. Note that in our setups the topological charge $Q$ of states generated in this way is given by \cite{nagaosa2013topological}
\begin{equation}\label{skyrmions:charge}
    Q = \frac{1}{4\pi}\int_{-\infty}^{\infty} \mathbf{m}\cdot (\partial_x\mathbf{m}\times\partial_y\mathbf{m}) ~dxdy = -k ~,
\end{equation}
which is therefore directly linked to the skyrmion (antiskyrmion) vorticity of $k =1$ ($k=-1$). After initialisation the skyrmions (antiskyrmions), the structure is optimised with respect to energy by the velocity projection optimization algorithm \cite{bessarab2015method}. From these configurations the radius is estimated according to Ref.~\cite{bocdanov1994properties}.

\noindent{{\textbf{Micromagnetic energies.}}}
Inserting the parametrization given by Eqs.~(\ref{eq:skyrmion_ansatz})-(\ref{eq:theta_profile}) into the micromagnetic energy functional~\cite{thiaville2012dynamics} yields the following expression for the exchange energy
\begin{equation}
    \begin{split}\label{eq:micromag_exchange}
        E_{\text{exc}} &= J \int_{-\infty}^{\infty}\left[ \left(\nabla_x\mathbf{m}\right)^2 + \left(\nabla_y\mathbf{m}\right)^2 \right]~d^2\mathbf{r} \\
        &= 2\pi J \left[\int_0^{\infty}\rho\left(\frac{\partial\Theta}{\partial\rho}\right)^2 + \frac{\sin^2(\Theta) |Q|^2}{\rho}~d\rho \right]
    \end{split}
\end{equation}
and similarly for the DMI energy 
\begin{equation}
    \begin{split}\label{eq:micromag_dmi}
        E_{\text{DM}} &= D \int_{-\infty}^{\infty}\left[m_z(\nabla\cdot\mathbf{m}) - (\mathbf{m}\cdot\nabla)m_z\right]~d^2\mathbf{r} \\
        &= 2\pi D \left[ \int_0^{\infty}\rho\frac{\partial \Theta}{\partial\rho} + \frac{\sin(2\Theta)Q}{2}~d\rho \right] \delta_{Q,-1} \cos\gamma
    \end{split}
\end{equation}
where $J$ and $D$ are the micromagnetic exchange and DMI constant, respectively. From Eq.~(\ref{eq:micromag_exchange}) it is evident that skyrmions with a topological charge of $Q=-1$ and antiskyrmions with $Q=+1$ are degenerate with respect to the exchange energy. In contrast, Eq.~(\ref{eq:micromag_dmi}) demonstrates that only skyrmions are affected by the DMI and that $E_{\rm DM}$ vanishes for any other value of the topological charge $Q$ due to the Kronecker delta term, $\delta_{Q,-1}$. Note, that the DMI energy also depends on the helicity $\gamma$ which prefers N\'eel-type skyrmions, i.e.~$\gamma =0$ (clockwise rotational sense) or $\gamma =\pi$ (counter clockwise rotational sense) depending on the sign of $D$.
    
\noindent{{\textbf{Computation of transition rates.}}}
The minimum energy path for a transition between two states A,B is computed using the geodesic nudged elastic band (GNEB) method \cite{bessarab2015method}. The highest energy point along the minimum energy path gives the SP, which is needed for the calculation of the transition rates $\Gamma^{\text{A}\to\text{B}}$ within the harmonic transition state theory~\cite{bessarab2012harmonic, goerzen2022atomistic}
\begin{equation}\label{eq:arrhenius_law}
    \Gamma^{\text{A}\to\text{B}} = \Gamma^{\text{A}\to\text{B}}_0\exp(-\beta\Delta E^{\text{A}\to\text{B}}),
\end{equation}
with $\beta = (k_BT)^{-1}$. Here, the activation energy $\Delta E^{\text{A}\to\text{B}} = E^{\text{SP}} -E^{\text{A}}$ is obtained as the energy difference between the initial state A and the transition-state SP, and the pre-exponential factor $\Gamma_0^{\text{A}\to\text{B}}$ takes into account the entropic contributions to the transition \cite{desplat2018thermal,varentcova2020toward,Malottki2019} and is given by \cite{bessarab2018lifetime, goerzen2022atomistic}
\begin{equation}
    \Gamma^{\text{A}\to\text{B}}_0 = \frac{v}{\sqrt{2\pi\beta}}\frac{\prod_{n=2}^{2N} Z_n^{\text{SP}}}{\prod_{n=1}^{2N}Z_n^{\text{A}}} ~.
\end{equation}
Here $v$ denotes the dynamical factor and $Z_n^X, X=\text{A, SP}$ are the partition functions of the initial and the transition states, which can either be treated in harmonic or zero mode approximation
\begin{equation}\label{eq:partition_functions}
    \arraycolsep=1.4pt\def\arraystretch{2.2}
    Z_n^X = \left\{\begin{array}{c}
    \displaystyle\sqrt{\frac{2\pi}{\beta\Omega_n^X}},\quad \Omega_n^X > 0, \\
    \displaystyle L_n^X,\quad \Omega_n^X = 0,
    \end{array}\right.
\end{equation}
where $\Omega_n^X$ are Hessian eigenvalues and $L_n^X$ denotes the length of the zero mode in space of spin configurations. Here only the harmonic approximation depends on temperature. This leads to a temperature dependence of $\Gamma_0\propto T^{{\Delta G}/2}$ for the pre-exponential factor where $\Delta G= G^{\text{X}}-G^{\text{SP}}$ is the difference in zero modes at the initial and saddle point state \cite{bessarab2018lifetime}. However the numerical treatment of the skyrmions helicity mode makes the temperature dependence even more complicated. The resulting temperature dependencies of the pre-exponential factors are summarised in Tab.~\ref{tab:number_zero_modes}. In this work we will deal with three different zero modes, which are addressed in detail in the following sections. 

\begin{table}[]
    \centering
    \renewcommand{\arraystretch}{2}
    \begin{tabularx}{0.45\textwidth}{|*{6}{Y|}}
        \hline
        $A\downarrow$ / $B\rightarrow$ & FM & Sk & Ask \\\hline
        FM & $-$ & $\frac{Z^{\ddagger}_h}{\sqrt{T}}$ $\left(\frac{1}{\sqrt{T}}\right)$ & $\frac{1}{\sqrt{T}}$\\\hline
        Sk & $T\frac{Z^{\ddagger}_h}{Z^{\text{Sk}}_h}$ $\left(\sqrt{T}\right)$ & $-$ & $-$ \\\hline
        Ask & $T$ & $-$ & $-$ \\\hline
    \end{tabularx}
    \caption{\textbf{Temperature dependence of pre-exponential factors.} The temperature dependence is given by $\Gamma_0^{A\to B}\propto T^{\Delta G/2}$ with $\Delta G$ being the difference of zero modes at the initial and transition state. Additionally the nonlinear temperature dependence induced by the full treatment of the skyrmions helicity mode is denoted by the partition function $Z^{\text{X}}_h=\int e^{-\beta E(\gamma)}d\gamma$ with $\beta=\left(k_BT\right)^{-1}$. The dependence is given for radial transitions between the FM, Sk, and Ask state. For skyrmions the terms in brackets denote the temperature dependence for Chimera transitions. The special case of skyrmion nucleation at $\eta=0$ (not listed in the table) has a dependence of $T^{-\frac{1}{2}}$. Dashes indicate absence of a transition between the corresponding states.}
    \label{tab:number_zero_modes}
\end{table}

\noindent{{\textbf{Partition function for zero modes.}}}
Zero modes \cite{braun1994fluctuations, watanabe2020counting} describe deformations of spin configurations that leave the energy of the state unchanged. Therefore, the curvature of the energy surface along this mode is zero, reflected by a vanishing Hessian eigenvalue of $\Omega\approx 0$ for the respective zero mode. A Boltzmann partition function $Z^X$ for such a mode becomes the length of a curve $\mathcal{C}$,
\begin{equation}\label{eq:zeromode_partitionfunction}
    Z^X = \int_{{\cal C}} e^{-\frac{\beta}{2}\Omega q^2}dq \approx \int_{{\cal C}}dq \ =L^X,
\end{equation}
where $\mathcal{C}$ describes the modes trajectory in configuration space.

\noindent{{\textbf{Translation zero mode.}}}
Translation zero modes describe in-plane displacements of a localised spin structure, such as the skyrmion. In our two dimensional film we therefore have two translation modes. To calculate the partition function for one of these modes the spin structure $\mathbf{M}(\mathbf{r})=(\mathbf{m}(\mathbf{r_1}), ...,\mathbf{m}(\mathbf{r_N}))$ is translated by an in-plane vector $\mathbf{a}$ which gives $\mathbf{M}(\mathbf{r}+\mathbf{a})$ The line integral from Eq.~(\ref{eq:zeromode_partitionfunction}) can then be evaluated using the geodesic distance between starting and end point of the curve \cite{varentcova2020toward}, reading
\begin{equation}
    Z_{\text{tr}} = \sqrt{\sum_{n=1}^N\norm{\mathbf{m}(\mathbf{r}_n+\mathbf{a})- \mathbf{m}(\mathbf{r}_n)}^2} 
\end{equation}
where the geodesic norm $\norm{\cdot}$ is the shortest distance between two points on the unit sphere.
    
\noindent{{\textbf{Helicity zero mode.}}}
Let now ${\cal C}$ be a closed curve associated with the helicity change as illustrated in Fig.~\ref{fig:helicity_concept}. In order to obtain an analytic form for the line integral from Eq.~(\ref{eq:zeromode_partitionfunction}), we parameterise the curve using the helicity $\gamma\in[0,2\pi]$
\begin{equation}\label{eq:zeromode_partitionfunction_helicity}
    Z_{\text{hl}} = \oint_{\mathcal{C}}dq = \int_{0}^{2\pi} \norm{\dot{\mathcal{C}}(\gamma)}d\gamma
\end{equation}
where the absolute of the gradient field is defined as $\norm{\dot{\mathcal{C}}(\gamma)} = \sqrt{\sum_{n=1}^N \left| \frac{d\mathbf{m}}{d\gamma}(\mathbf{r}_n,\gamma) \right|^2 }$. We now observe, that following the helicity mode, every spin $\mathbf{m}(\mathbf{r}_n)$ performs a full rotation in the xy-plane. On the unit sphere it therefore travels on a ring with length $L_n= 2\pi\sqrt{\left(m^x(\mathbf{r}_n)\right)^2+\left(m^y(\mathbf{r}_n)\right)^2}= 2\pi\sqrt{1-\left(m^z(\mathbf{r}_n)\right)^2}$. Since this rotation is homogeneous, the absolute of the spins gradient along the curve becomes constant with $\left|\frac{d\mathbf{m}}{d\gamma}(\mathbf{r}_n,\gamma)\right|= \frac{L_n}{\gamma_1-\gamma_0}=\sqrt{1-\left(m^z(\mathbf{r}_n)\right)^2}$ where $\gamma_0=0$ and $\gamma_1=2\pi$ are the integration limits. Inserting this into Eq.~(\ref{eq:zeromode_partitionfunction_helicity}) we obtain the analytic form for the helicity modes partition function in zero mode approximation
\begin{equation}\label{eq:partfunc_helicity}
    Z_{\gamma} = 2\pi\sqrt{\sum_{n=1}^N 1-\left(m^z(\mathbf{r}_n)\right)^2}
\end{equation}
In the case of skyrmions the helicity mode cannot be treated as a zero mode anymore in the presence of DMI. We therefore model the homogeneous rotation of spins along the mode by a series of states $\mathbf{M}_k$, where spins are iteratively rotated around the $z$ axis by the change in helicity $\Delta\gamma$
\begin{equation}
    \mathbf{m}_{k}(\mathbf{r}) = \mathbf{R}^z_{\Delta\gamma}\mathbf{m}_{k-1}\left(\mathbf{r}\right)~,
\end{equation}
From this series of images the Boltzmann partition function can be numerically computed
\begin{equation}\label{eq:partfunc_num}
    Z = \sum_{k=1}^{K} e^{-\beta (E_k-E_0)} \sqrt{\sum_{n=1}^N \norm{ \mathbf{m}_k(\mathbf{r}_n) - \mathbf{m}_{k-1}(\mathbf{r}_n) }^2}
\end{equation}
    
\noindent{{\textbf{Chimera rotation zero mode.}}}
Another zero mode that occurs in our simulations is the rotation mode of the chimera SP, describing a full rotation of the structure around the Bloch point (cf. Ref.~\cite{Muckel2021}). To our knowledge,  no analytic expression for the partition function of this particular mode has been presented yet. We compute the corresponding Boltzmann partition function numerically as in Eq.~(\ref{eq:partfunc_num}). For that we parameterise the images along the mode by the angle $\varphi$ so that spins at the same position $\mathbf{r}_n$ but in consecutive images $k-1$ and $k$ read \cite{kuchkin2021geometry}
\begin{equation}
    \mathbf{m}_{k}(\mathbf{r}_n -\mathbf{c}) = \mathbf{R}^z_{\Delta\varphi}\mathbf{m}_{k-1}\left(\mathbf{R}^z_{-\Delta\varphi}[\mathbf{r}_n -\mathbf{c}]\right)~.
\end{equation}
Here $\mathbf{c}$ is the center of rotation on the lattice, which has been chosen to be the location of the saddle points Bloch-point-like configuration. This position is computed as the point where the absolute of the states interpolated magnetisation vanishes, therefore $\mathbf{c}=\argmin_{\mathbf{r}}|\mathbf{m}(\mathbf{r})|$.

\noindent{{\textbf{Master equations.}}}
Consider a system, that can be in three different states: the skyrmion state, the antiskyrmion state, and the FM state. Based on the calculated transition rates $\Gamma^{\text{m}\to\text{n}}$ between those states, we can simulate the time evolution of the probabilities of the states, as described in the following. Let $p^\text{A}$ be the time-dependent probability for the system to be in a specific state A = FM, Sk, Ask. We construct the vector
 \begin{equation}
    \mathbf{p}(t)= \left(p^{\text{FM}}(t), p^{\text{Sk}}(t),p^{\text{Ask}}(t)\right)^T,\quad \sum_{\text{A}} p^{\text{A}}(t)=1
\end{equation}
The equation describing the evolution of $\mathbf{p}(t)$ is the master equation
\begin{equation}
    \frac{d}{dt}\mathbf{p}(t) = \mathbf{U}\mathbf{p}(t)
\end{equation}
with transition matrix $\mathbf{U}$
\begin{equation}
    \left(\mathbf{U}\right)_{AB}=\left\{ \begin{array}{c}
         \Gamma^{B\to A} \quad\text{if}\quad \text{A}\neq \text{B}  \\
         -\sum_{A} \Gamma^{B\to A} \quad\text{else}
    \end{array} \right.
\end{equation}
The general solution to the master equation can then be computed from the three eigenpairs $(\lambda_n, \mathbf{v}_n)$ of $\mathbf{U}$
\begin{equation}
    \bm{p}(t) = \sum_{n=1}^3 c_ne^{\lambda_nt}\mathbf{v}_n,\quad \mathbf{U}\mathbf{v}_n = \lambda_n \mathbf{v}_n
\end{equation}
In the last step, the constants $c_n$ have to be calculated from the initial conditions which, in our case, correspond to the initialization of the system in the FM state. The equilibrium distribution $\mathbf{p}_{\infty}$ for large times $t\to\infty$, which fulfills
\begin{equation}
    0=\mathbf{U}\mathbf{p}_{\infty}
\end{equation}
is directly computed from the nullspace of $\mathbf{U}$.

The temperature dependence of the solutions stems from the temperature dependence of the rates themselves (cf. Tab.~\ref{tab:number_zero_modes}). In the setup of the transition matrix we also take into account that the probability for finding a skyrmion or antiskyrmion increases with the search area. This is done by scaling only the nucleation rates by the number of possible nucleation saddle points in a given area $F$, reading on the hexagonal lattice $2F = a^2\sin(\pi/3)N_{sp}$ with the Iridium lattice constant of $a=0.2715$~nm. Physically this reflects the fact that a nucleation can take place at any interstitial site of the lattice \cite{goerzen2022atomistic, Muckel2021}. In this regard, the probabilities in the inset of Fig.~\ref{fig:master_equation}b show the probability for finding at least one skyrmion, antiskyrmion or none of both (FM) in an area $F=1~\mu$m$^2$. Note that the ratio of skyrmion and antiskyrmion probability has been found to be independent of $F$. 

\section*{}
\noindent{{\large\textbf{Data availability}.}\newline The data presented in this paper are available from the authors upon reasonable request.

\section*{}
\noindent{{\large\textbf{Code availability}.}\newline The atomistic spin dynamics code is available from the authors upon reasonable request.

\bibliographystyle{naturemag}

\begin{thebibliography}{10}
\expandafter\ifx\csname url\endcsname\relax
  \def\url#1{\texttt{#1}}\fi
\expandafter\ifx\csname urlprefix\endcsname\relax\def\urlprefix{URL }\fi
\providecommand{\bibinfo}[2]{#2}
\providecommand{\eprint}[2][]{\url{#2}}

\bibitem{Muehlbauer2009}
\bibinfo{author}{M{\"u}hlbauer, S.} \emph{et~al.}
\newblock \bibinfo{title}{Skyrmion lattice in a chiral magnet}.
\newblock \emph{\bibinfo{journal}{Science}} \textbf{\bibinfo{volume}{323}},
  \bibinfo{pages}{915--919} (\bibinfo{year}{2009}).

\bibitem{Yu2010}
\bibinfo{author}{Yu, X.} \emph{et~al.}
\newblock \bibinfo{title}{Real-space observation of a two-dimensional skyrmion
  crystal}.
\newblock \emph{\bibinfo{journal}{Nature}} \textbf{\bibinfo{volume}{465}},
  \bibinfo{pages}{901--904} (\bibinfo{year}{2010}).

\bibitem{Heinze2011}
\bibinfo{author}{Heinze, S.} \emph{et~al.}
\newblock \bibinfo{title}{Spontaneous atomic-scale magnetic skyrmion lattice in
  two dimensions}.
\newblock \emph{\bibinfo{journal}{Nat. Phys.}} \textbf{\bibinfo{volume}{7}},
  \bibinfo{pages}{713} (\bibinfo{year}{2011}).

\bibitem{Romming2013}
\bibinfo{author}{Romming, N.} \emph{et~al.}
\newblock \bibinfo{title}{Writing and deleting single magnetic skyrmions}.
\newblock \emph{\bibinfo{journal}{Science}} \textbf{\bibinfo{volume}{341}},
  \bibinfo{pages}{636--639} (\bibinfo{year}{2013}).

\bibitem{nagaosa2013topological}
\bibinfo{author}{Nagaosa, N.} \& \bibinfo{author}{Tokura, Y.}
\newblock \bibinfo{title}{Topological properties and dynamics of magnetic
  skyrmions}.
\newblock \emph{\bibinfo{journal}{Nat. Nanotechnol.}}
  \textbf{\bibinfo{volume}{8}}, \bibinfo{pages}{899--911}
  (\bibinfo{year}{2013}).

\bibitem{Fert2017}
\bibinfo{author}{Fert, A.}, \bibinfo{author}{Reyren, N.} \&
  \bibinfo{author}{Cros, V.}
\newblock \bibinfo{title}{Magnetic skyrmions: advances in physics and potential
  applications}.
\newblock \emph{\bibinfo{journal}{Nat. Rev. Mater.}}
  \textbf{\bibinfo{volume}{2}}, \bibinfo{pages}{17031} (\bibinfo{year}{2017}).

\bibitem{Fert2013}
\bibinfo{author}{Fert, A.}, \bibinfo{author}{Cros, V.} \&
  \bibinfo{author}{Sampaio, J.}
\newblock \bibinfo{title}{Skyrmions on the track}.
\newblock \emph{\bibinfo{journal}{Nat. Nanotechnol.}}
  \textbf{\bibinfo{volume}{8}}, \bibinfo{pages}{152--156}
  (\bibinfo{year}{2013}).

\bibitem{Song2020}
\bibinfo{author}{Song, K.~M.} \emph{et~al.}
\newblock \bibinfo{title}{Skyrmion-based artificial synapses for neuromorphic
  computing}.
\newblock \emph{\bibinfo{journal}{Nat. Electron.}}
  \textbf{\bibinfo{volume}{3}}, \bibinfo{pages}{148--155}
  (\bibinfo{year}{2020}).

\bibitem{pinna2018skyrmion}
\bibinfo{author}{Pinna, D.} \emph{et~al.}
\newblock \bibinfo{title}{Skyrmion gas manipulation for probabilistic
  computing}.
\newblock \emph{\bibinfo{journal}{Phys. Rev. Appl.}}
  \textbf{\bibinfo{volume}{9}}, \bibinfo{pages}{064018} (\bibinfo{year}{2018}).

\bibitem{grollier2020neuromorphic}
\bibinfo{author}{Grollier, J.} \emph{et~al.}
\newblock \bibinfo{title}{Neuromorphic spintronics}.
\newblock \emph{\bibinfo{journal}{Nat. Electron.}}
  \textbf{\bibinfo{volume}{3}}, \bibinfo{pages}{360--370}
  (\bibinfo{year}{2020}).

\bibitem{Psaroudaki2021}
\bibinfo{author}{Psaroudaki, C.} \& \bibinfo{author}{Panagopoulos, C.}
\newblock \bibinfo{title}{Skyrmion qubits: A new class of quantum logic
  elements based on nanoscale magnetization}.
\newblock \emph{\bibinfo{journal}{Phys. Rev. Lett.}}
  \textbf{\bibinfo{volume}{127}}, \bibinfo{pages}{067201}
  (\bibinfo{year}{2021}).

\bibitem{Romming2015}
\bibinfo{author}{Romming, N.}, \bibinfo{author}{Kubetzka, A.},
  \bibinfo{author}{Hanneken, C.}, \bibinfo{author}{von Bergmann, K.} \&
  \bibinfo{author}{Wiesendanger, R.}
\newblock \bibinfo{title}{Field-dependent size and shape of single magnetic
  skyrmions}.
\newblock \emph{\bibinfo{journal}{Phys. Rev. Lett.}}
  \textbf{\bibinfo{volume}{114}}, \bibinfo{pages}{177203}
  (\bibinfo{year}{2015}).

\bibitem{Herve2018}
\bibinfo{author}{Herv{\'{e}}, M.} \emph{et~al.}
\newblock \bibinfo{title}{Stabilizing spin spirals and isolated skyrmions at
  low magnetic field exploiting vanishing magnetic anisotropy}.
\newblock \emph{\bibinfo{journal}{Nat. Commun.}} \textbf{\bibinfo{volume}{9}},
  \bibinfo{pages}{1015} (\bibinfo{year}{2018}).

\bibitem{Meyer2019}
\bibinfo{author}{Meyer, S.} \emph{et~al.}
\newblock \bibinfo{title}{Isolated zero field sub-10 nm skyrmions in ultrathin
  co films}.
\newblock \emph{\bibinfo{journal}{Nat. Commun.}} \textbf{\bibinfo{volume}{10}},
  \bibinfo{pages}{3823} (\bibinfo{year}{2019}).

\bibitem{MoreauLuchaire2016}
\bibinfo{author}{Moreau-Luchaire, C.} \emph{et~al.}
\newblock \bibinfo{title}{Additive interfacial chiral interaction in
  multilayers for stabilization of small individual skyrmions at room
  temperature}.
\newblock \emph{\bibinfo{journal}{Nat. Nanotechnol.}}
  \textbf{\bibinfo{volume}{11}}, \bibinfo{pages}{444--448}
  (\bibinfo{year}{2016}).

\bibitem{Boulle2016}
\bibinfo{author}{Boulle, O.} \emph{et~al.}
\newblock \bibinfo{title}{Room-temperature chiral magnetic skyrmions in
  ultrathin magnetic nanostructures}.
\newblock \emph{\bibinfo{journal}{Nat. Nanotechnol.}}
  \textbf{\bibinfo{volume}{11}}, \bibinfo{pages}{449--454}
  (\bibinfo{year}{2016}).

\bibitem{Woo2016}
\bibinfo{author}{Woo, S.} \emph{et~al.}
\newblock \bibinfo{title}{Observation of room-temperature magnetic skyrmions
  and their current-driven dynamics in ultrathin metallic ferromagnets}.
\newblock \emph{\bibinfo{journal}{Nat. Mater.}} \textbf{\bibinfo{volume}{15}},
  \bibinfo{pages}{501--506} (\bibinfo{year}{2016}).

\bibitem{Soumyanarayanan2017}
\bibinfo{author}{Soumyanarayanan, A.} \emph{et~al.}
\newblock \bibinfo{title}{Tunable room-temperature magnetic skyrmions in
  {I}r/{F}e/{C}o/{P}t multilayers}.
\newblock \emph{\bibinfo{journal}{Nat. Mater.}} \textbf{\bibinfo{volume}{16}},
  \bibinfo{pages}{898--904} (\bibinfo{year}{2017}).

\bibitem{Jiang2017}
\bibinfo{author}{Jiang, W.} \emph{et~al.}
\newblock \bibinfo{title}{Direct observation of the skyrmion hall effect}.
\newblock \emph{\bibinfo{journal}{Nat. Phys.}} \textbf{\bibinfo{volume}{13}},
  \bibinfo{pages}{162--169} (\bibinfo{year}{2017}).

\bibitem{Litzius2017}
\bibinfo{author}{Litzius, K.} \emph{et~al.}
\newblock \bibinfo{title}{Skyrmion {H}all effect revealed by direct
  time-resolved {X}-ray microscopy}.
\newblock \emph{\bibinfo{journal}{Nat. Phys.}} \textbf{\bibinfo{volume}{13}},
  \bibinfo{pages}{170--175} (\bibinfo{year}{2017}).

\bibitem{Hoffmann2017}
\bibinfo{author}{Hoffmann, M.} \emph{et~al.}
\newblock \bibinfo{title}{Antiskyrmions stabilized at interfaces by anisotropic
  {D}zyaloshinskii-{M}oriya interactions}.
\newblock \emph{\bibinfo{journal}{Nat. Commun.}} \textbf{\bibinfo{volume}{8}},
  \bibinfo{pages}{308} (\bibinfo{year}{2017}).

\bibitem{Goebel2021}
\bibinfo{author}{G\"obel, B.}, \bibinfo{author}{Mertig, I.} \&
  \bibinfo{author}{Tretiakov, O.~A.}
\newblock \bibinfo{title}{Beyond skyrmions: Review and perspectives of
  alternative magnetic quasiparticles}.
\newblock \emph{\bibinfo{journal}{Phys. Rep.}} \textbf{\bibinfo{volume}{895}},
  \bibinfo{pages}{1--28} (\bibinfo{year}{2021}).

\bibitem{Leonov2015}
\bibinfo{author}{Leonov, A.~O.} \& \bibinfo{author}{Mostovoy, M.}
\newblock \bibinfo{title}{Multiply periodic states and isolated skyrmions in an
  anisotropic frustrated magnet}.
\newblock \emph{\bibinfo{journal}{Nat. Commun.}} \textbf{\bibinfo{volume}{6}},
  \bibinfo{pages}{8275} (\bibinfo{year}{2015}).

\bibitem{Rozsa2017}
\bibinfo{author}{R{\'{o}}zsa, L.} \emph{et~al.}
\newblock \bibinfo{title}{Formation and stability of metastable skyrmionic spin
  structures with various topologies in an ultrathin film}.
\newblock \emph{\bibinfo{journal}{Phys. Rev. B}} \textbf{\bibinfo{volume}{95}},
  \bibinfo{pages}{094423} (\bibinfo{year}{2017}).

\bibitem{Goebel2019}
\bibinfo{author}{G\"obel, B.}, \bibinfo{author}{Mook, A.},
  \bibinfo{author}{Henk, J.}, \bibinfo{author}{Mertig, I.} \&
  \bibinfo{author}{Tretiakov, O.~A.}
\newblock \bibinfo{title}{Magnetic bimerons as skyrmion analogues in in-plane
  magnets}.
\newblock \emph{\bibinfo{journal}{Phys. Rev. B}} \textbf{\bibinfo{volume}{99}},
  \bibinfo{pages}{060407} (\bibinfo{year}{2019}).

\bibitem{Rybakov2019}
\bibinfo{author}{Rybakov, F.~N.} \& \bibinfo{author}{Kiselev, N.~S.}
\newblock \bibinfo{title}{Chiral magnetic skyrmions with arbitrary topological
  charge}.
\newblock \emph{\bibinfo{journal}{Phys. Rev. B}} \textbf{\bibinfo{volume}{99}},
  \bibinfo{pages}{064437} (\bibinfo{year}{2019}).

\bibitem{Kuchkin2020}
\bibinfo{author}{Kuchkin, V.~M.} \emph{et~al.}
\newblock \bibinfo{title}{Magnetic skyrmions, chiral kinks, and holomorphic
  functions}.
\newblock \emph{\bibinfo{journal}{Phys. Rev. B}}
  \textbf{\bibinfo{volume}{102}}, \bibinfo{pages}{144422}
  (\bibinfo{year}{2020}).

\bibitem{Nayak2017}
\bibinfo{author}{Nayak, A.~K.} \emph{et~al.}
\newblock \bibinfo{title}{Magnetic antiskyrmions above room temperature in
  tetragonal {H}eusler materials}.
\newblock \emph{\bibinfo{journal}{Nature}} \textbf{\bibinfo{volume}{548}},
  \bibinfo{pages}{561--566} (\bibinfo{year}{2017}).

\bibitem{Potkina2020}
\bibinfo{author}{Potkina, M.~N.}, \bibinfo{author}{Lobanov, I.~S.},
  \bibinfo{author}{Tretiakov, O.~A.}, \bibinfo{author}{J\'onsson, H.} \&
  \bibinfo{author}{Uzdin, V.~M.}
\newblock \bibinfo{title}{Stability of long-lived antiskyrmions in the
  {M}n-{P}t-{S}n tetragonal {H}eusler material}.
\newblock \emph{\bibinfo{journal}{Phys. Rev. B}}
  \textbf{\bibinfo{volume}{102}}, \bibinfo{pages}{134430}
  (\bibinfo{year}{2020}).

\bibitem{Ga2022}
\bibinfo{author}{Ga, Y.} \emph{et~al.}
\newblock \bibinfo{title}{Anisotropic {D}zyaloshinskii-{M}oriya interaction
  protected by {D$_{\rm 2d}$} crystal symmetry in two-dimensional ternary
  compounds}.
\newblock \emph{\bibinfo{journal}{npj Comp. Mater.}}
  \textbf{\bibinfo{volume}{8}}, \bibinfo{pages}{128} (\bibinfo{year}{2022}).

\bibitem{Jena2020}
\bibinfo{author}{Jena, J.} \emph{et~al.}
\newblock \bibinfo{title}{Elliptical {B}loch skyrmion chiral twins in an
  antiskyrmion system}.
\newblock \emph{\bibinfo{journal}{Nat. Commun.}} \textbf{\bibinfo{volume}{11}},
  \bibinfo{pages}{1115} (\bibinfo{year}{2020}).

\bibitem{Peng2020}
\bibinfo{author}{Peng, L.} \emph{et~al.}
\newblock \bibinfo{title}{Controlled transformation of skyrmions and
  antiskyrmions in a non-centrosymmetric magnet}.
\newblock \emph{\bibinfo{journal}{Nat. Nanotechnol.}}
  \textbf{\bibinfo{volume}{15}}, \bibinfo{pages}{181--186}
  (\bibinfo{year}{2020}).

\bibitem{karube2022high}
\bibinfo{author}{Karube, K.} \& \bibinfo{author}{Taguchi, Y.}
\newblock \bibinfo{title}{High-temperature non-centrosymmetric magnets for
  skyrmionics}.
\newblock \emph{\bibinfo{journal}{APL Mater.}} \textbf{\bibinfo{volume}{10}},
  \bibinfo{pages}{080902} (\bibinfo{year}{2022}).

\bibitem{Koshibae2016}
\bibinfo{author}{Koshibae, W.} \& \bibinfo{author}{Nagaosa, N.}
\newblock \bibinfo{title}{Theory of antiskyrmions in magnets}.
\newblock \emph{\bibinfo{journal}{Nat. Commun.}} \textbf{\bibinfo{volume}{7}},
  \bibinfo{pages}{10542} (\bibinfo{year}{2016}).

\bibitem{Heigl2021}
\bibinfo{author}{Heigl, M.} \emph{et~al.}
\newblock \bibinfo{title}{Dipolar-stabilized first and second-order
  antiskyrmions in ferrimagnetic multilayers}.
\newblock \emph{\bibinfo{journal}{Nat. Commun.}} \textbf{\bibinfo{volume}{12}},
  \bibinfo{pages}{2611} (\bibinfo{year}{2021}).

\bibitem{Kuchkin2020a}
\bibinfo{author}{Kuchkin, V.~M.} \& \bibinfo{author}{Kiselev, N.~S.}
\newblock \bibinfo{title}{Turning a chiral skyrmion inside out}.
\newblock \emph{\bibinfo{journal}{Phys. Rev. B}}
  \textbf{\bibinfo{volume}{101}}, \bibinfo{pages}{064408}
  (\bibinfo{year}{2020}).

\bibitem{Zheng2022}
\bibinfo{author}{Zheng, F.} \emph{et~al.}
\newblock \bibinfo{title}{Skyrmion–antiskyrmion pair creation and
  annihilation in a cubic chiral magnet}.
\newblock \emph{\bibinfo{journal}{Nat. Phys.}} \textbf{\bibinfo{volume}{18}},
  \bibinfo{pages}{863--868} (\bibinfo{year}{2022}).

\bibitem{Lin2016}
\bibinfo{author}{Lin, S.-Z.} \& \bibinfo{author}{Hayami, S.}
\newblock \bibinfo{title}{Ginzburg-{L}andau theory for skyrmions in
  inversion-symmetric magnets with competing interactions}.
\newblock \emph{\bibinfo{journal}{Phys. Rev. B}} \textbf{\bibinfo{volume}{93}},
  \bibinfo{pages}{064430} (\bibinfo{year}{2016}).

\bibitem{Malottki2017}
\bibinfo{author}{von Malottki, S.}, \bibinfo{author}{Dup{\'e}, B.},
  \bibinfo{author}{Bessarab, P.}, \bibinfo{author}{Delin, A.} \&
  \bibinfo{author}{Heinze, S.}
\newblock \bibinfo{title}{Enhanced skyrmion stability due to exchange
  frustration}.
\newblock \emph{\bibinfo{journal}{Sci. Rep.}} \textbf{\bibinfo{volume}{7}},
  \bibinfo{pages}{12299} (\bibinfo{year}{2017}).

\bibitem{Zhang2017}
\bibinfo{author}{Zhang, X.} \emph{et~al.}
\newblock \bibinfo{title}{Skyrmion dynamics in a frustrated ferromagnetic film
  and current-induced helicity locking-unlocking transition}.
\newblock \emph{\bibinfo{journal}{Nat. Commun.}} \textbf{\bibinfo{volume}{8}},
  \bibinfo{pages}{1717} (\bibinfo{year}{2017}).

\bibitem{Desplat2019}
\bibinfo{author}{Desplat, L.}, \bibinfo{author}{Kim, J.-V.} \&
  \bibinfo{author}{Stamps, R.~L.}
\newblock \bibinfo{title}{Paths to annihilation of first- and second-order
  (anti)skyrmions via (anti)meron nucleation on the frustrated square lattice}.
\newblock \emph{\bibinfo{journal}{Phys. Rev. B}} \textbf{\bibinfo{volume}{99}},
  \bibinfo{pages}{174409} (\bibinfo{year}{2019}).

\bibitem{Paul2020}
\bibinfo{author}{Paul, S.}, \bibinfo{author}{Haldar, S.}, \bibinfo{author}{von
  Malottki, S.} \& \bibinfo{author}{Heinze, S.}
\newblock \bibinfo{title}{{Role of higher-order exchange interactions for
  skyrmion stability}}.
\newblock \emph{\bibinfo{journal}{Nat. Commun.}} \textbf{\bibinfo{volume}{11}},
  \bibinfo{pages}{4756} (\bibinfo{year}{2020}).

\bibitem{Perini2019}
\bibinfo{author}{Perini, M.} \emph{et~al.}
\newblock \bibinfo{title}{Electrical detection of domain walls and skyrmions in
  co films using noncollinear magnetoresistance}.
\newblock \emph{\bibinfo{journal}{Phys. Rev. Lett.}}
  \textbf{\bibinfo{volume}{123}}, \bibinfo{pages}{237205}
  (\bibinfo{year}{2019}).

\bibitem{dupe2014tailoring}
\bibinfo{author}{Dup{\'e}, B.}, \bibinfo{author}{Hoffmann, M.},
  \bibinfo{author}{Paillard, C.} \& \bibinfo{author}{Heinze, S.}
\newblock \bibinfo{title}{Tailoring magnetic skyrmions in ultra-thin transition
  metal films}.
\newblock \emph{\bibinfo{journal}{Nat. Commun.}} \textbf{\bibinfo{volume}{5}},
  \bibinfo{pages}{4030} (\bibinfo{year}{2014}).

\bibitem{Hagemeister2015}
\bibinfo{author}{Hagemeister, J.}, \bibinfo{author}{Romming, N.},
  \bibinfo{author}{Von~Bergmann, K.}, \bibinfo{author}{Vedmedenko, E.} \&
  \bibinfo{author}{Wiesendanger, R.}
\newblock \bibinfo{title}{Stability of single skyrmionic bits}.
\newblock \emph{\bibinfo{journal}{Nat. Commun.}} \textbf{\bibinfo{volume}{6}},
  \bibinfo{pages}{8455} (\bibinfo{year}{2015}).

\bibitem{Malottki2019}
\bibinfo{author}{von Malottki, S.}, \bibinfo{author}{Bessarab, P.~F.},
  \bibinfo{author}{Haldar, S.}, \bibinfo{author}{Delin, A.} \&
  \bibinfo{author}{Heinze, S.}
\newblock \bibinfo{title}{Skyrmion lifetime in ultrathin films}.
\newblock \emph{\bibinfo{journal}{Phys. Rev. B}} \textbf{\bibinfo{volume}{99}},
  \bibinfo{pages}{060409} (\bibinfo{year}{2019}).

\bibitem{Muckel2021}
\bibinfo{author}{Muckel, F.} \emph{et~al.}
\newblock \bibinfo{title}{Experimental identification of two distinct skyrmion
  collapse mechanisms}.
\newblock \emph{\bibinfo{journal}{Nat. Phys.}} \textbf{\bibinfo{volume}{17}},
  \bibinfo{pages}{395} (\bibinfo{year}{2021}).

\bibitem{bocdanov1994properties}
\bibinfo{author}{Bocdanov, A.} \& \bibinfo{author}{Hubert, A.}
\newblock \bibinfo{title}{The properties of isolated magnetic vortices}.
\newblock \emph{\bibinfo{journal}{Phys. Status Solidi (b)}}
  \textbf{\bibinfo{volume}{186}}, \bibinfo{pages}{527--543}
  (\bibinfo{year}{1994}).

\bibitem{bessarab2015method}
\bibinfo{author}{Bessarab, P.~F.}, \bibinfo{author}{Uzdin, V.~M.} \&
  \bibinfo{author}{J{\'o}nsson, H.}
\newblock \bibinfo{title}{Method for finding mechanism and activation energy of
  magnetic transitions, applied to skyrmion and antivortex annihilation}.
\newblock \emph{\bibinfo{journal}{Comput. Phys. Commun.}}
  \textbf{\bibinfo{volume}{196}}, \bibinfo{pages}{335--347}
  (\bibinfo{year}{2015}).

\bibitem{Lindner2020}
\bibinfo{author}{Lindner, P.} \emph{et~al.}
\newblock \bibinfo{title}{Temperature and magnetic field dependent behavior of
  atomic-scale skyrmions in {P}d/{F}e/{I}r(111) nanoislands}.
\newblock \emph{\bibinfo{journal}{Phys. Rev. B}}
  \textbf{\bibinfo{volume}{101}}, \bibinfo{pages}{214445}
  (\bibinfo{year}{2020}).

\bibitem{Boettcher2018}
\bibinfo{author}{B{\"o}ttcher, M.}, \bibinfo{author}{Heinze, S.},
  \bibinfo{author}{Egorov, S.}, \bibinfo{author}{Sinova, J.} \&
  \bibinfo{author}{Dup{\'e}, B.}
\newblock \bibinfo{title}{B--{T} phase diagram of {P}d/{F}e/{I}r (111) computed
  with parallel tempering {M}onte {C}arlo}.
\newblock \emph{\bibinfo{journal}{New J. Phys.}} \textbf{\bibinfo{volume}{20}},
  \bibinfo{pages}{103014} (\bibinfo{year}{2018}).

\bibitem{bessarab2018lifetime}
\bibinfo{author}{Bessarab, P.~F.} \emph{et~al.}
\newblock \bibinfo{title}{Lifetime of racetrack skyrmions}.
\newblock \emph{\bibinfo{journal}{Sci. Rep.}} \textbf{\bibinfo{volume}{8}},
  \bibinfo{pages}{3433} (\bibinfo{year}{2018}).

\bibitem{goerzen2022atomistic}
\bibinfo{author}{Goerzen, M.~A.}, \bibinfo{author}{von Malottki, S.},
  \bibinfo{author}{Kwiatkowski, G.~J.}, \bibinfo{author}{Bessarab, P.~F.} \&
  \bibinfo{author}{Heinze, S.}
\newblock \bibinfo{title}{Atomistic spin simulations of electric-field-assisted
  nucleation and annihilation of magnetic skyrmions in {P}d/{F}e/{I}r(111)}.
\newblock \emph{\bibinfo{journal}{Phys. Rev. B}}
  \textbf{\bibinfo{volume}{105}}, \bibinfo{pages}{214435}
  (\bibinfo{year}{2022}).

\bibitem{desplat2018thermal}
\bibinfo{author}{Desplat, L.}, \bibinfo{author}{Suess, D.},
  \bibinfo{author}{Kim, J.-V.} \& \bibinfo{author}{Stamps, R.}
\newblock \bibinfo{title}{Thermal stability of metastable magnetic skyrmions:
  Entropic narrowing and significance of internal eigenmodes}.
\newblock \emph{\bibinfo{journal}{Phys. Rev. B}} \textbf{\bibinfo{volume}{98}},
  \bibinfo{pages}{134407} (\bibinfo{year}{2018}).

\bibitem{varentcova2020toward}
\bibinfo{author}{Varentcova, A.~S.} \emph{et~al.}
\newblock \bibinfo{title}{Toward room-temperature nanoscale skyrmions in
  ultrathin films}.
\newblock \emph{\bibinfo{journal}{npj Comput. Mater.}}
  \textbf{\bibinfo{volume}{6}}, \bibinfo{pages}{193} (\bibinfo{year}{2020}).

\bibitem{potkina2020skyrmions}
\bibinfo{author}{Potkina, M.~N.}, \bibinfo{author}{Lobanov, I.~S.},
  \bibinfo{author}{J{\'o}nsson, H.} \& \bibinfo{author}{Uzdin, V.~M.}
\newblock \bibinfo{title}{Skyrmions in antiferromagnets: Thermal stability and
  the effect of external field and impurities}.
\newblock \emph{\bibinfo{journal}{J. Appl. Phys.}}
  \textbf{\bibinfo{volume}{127}}, \bibinfo{pages}{213906}
  (\bibinfo{year}{2020}).

\bibitem{schrautzer2022effects}
\bibinfo{author}{Schrautzer, H.}, \bibinfo{author}{von Malottki, S.},
  \bibinfo{author}{Bessarab, P.~F.} \& \bibinfo{author}{Heinze, S.}
\newblock \bibinfo{title}{Effects of interlayer exchange on collapse mechanisms
  and stability of magnetic skyrmions}.
\newblock \emph{\bibinfo{journal}{Phys. Rev. B}}
  \textbf{\bibinfo{volume}{105}}, \bibinfo{pages}{014414}
  (\bibinfo{year}{2022}).

\bibitem{desplat2020entropy}
\bibinfo{author}{Desplat, L.} \& \bibinfo{author}{Kim, J.-V.}
\newblock \bibinfo{title}{Entropy-reduced retention times in magnetic memory
  elements: A case of the {M}eyer-{N}eldel compensation rule}.
\newblock \emph{\bibinfo{journal}{Phys. Rev. Lett.}}
  \textbf{\bibinfo{volume}{125}}, \bibinfo{pages}{107201}
  (\bibinfo{year}{2020}).

\bibitem{haldar2018first}
\bibinfo{author}{Haldar, S.}, \bibinfo{author}{von Malottki, S.},
  \bibinfo{author}{Meyer, S.}, \bibinfo{author}{Bessarab, P.~F.} \&
  \bibinfo{author}{Heinze, S.}
\newblock \bibinfo{title}{First-principles prediction of sub-10-nm skyrmions in
  {P}d/{F}e bilayers on {R}h(111)}.
\newblock \emph{\bibinfo{journal}{Phys. Rev. B}} \textbf{\bibinfo{volume}{98}},
  \bibinfo{pages}{060413} (\bibinfo{year}{2018}).

\bibitem{zurek1985cosmological}
\bibinfo{author}{Zurek, W.~H.}
\newblock \bibinfo{title}{Cosmological experiments in superfluid helium?}
\newblock \emph{\bibinfo{journal}{Nature}} \textbf{\bibinfo{volume}{317}},
  \bibinfo{pages}{505--508} (\bibinfo{year}{1985}).

\bibitem{kibble1976topology}
\bibinfo{author}{Kibble, T.~W.}
\newblock \bibinfo{title}{Topology of cosmic domains and strings}.
\newblock \emph{\bibinfo{journal}{J. Phys. A: Math. Gen.}}
  \textbf{\bibinfo{volume}{9}}, \bibinfo{pages}{1387} (\bibinfo{year}{1976}).

\bibitem{Li2022}
\bibinfo{author}{Li, D.}, \bibinfo{author}{Haldar, S.} \&
  \bibinfo{author}{Heinze, S.}
\newblock \bibinfo{title}{Strain-driven zero-field near-10 nm skyrmions in
  two-dimensional van der {W}aals heterostructures}.
\newblock \emph{\bibinfo{journal}{Nano Lett.}} \textbf{\bibinfo{volume}{22}},
  \bibinfo{pages}{7706--7713} (\bibinfo{year}{2022}).

\bibitem{Nickel2023}
\bibinfo{author}{Nickel, F.}, \bibinfo{author}{Meyer, S.} \&
  \bibinfo{author}{Heinze, S.}
\newblock \bibinfo{title}{Exchange and {D}zyaloshinskii-{M}oriya interaction in
  {R}h/{C}o/{F}e/{I}r multilayers: towards skyrmions in exchange-frustrated
  multilayers}.
\newblock \emph{\bibinfo{journal}{appears in Phys. Rev. B}}
  (\bibinfo{year}{2023}).

\bibitem{thiaville2012dynamics}
\bibinfo{author}{Thiaville, A.}, \bibinfo{author}{Rohart, S.},
  \bibinfo{author}{Ju{\'e}, {\'E}.}, \bibinfo{author}{Cros, V.} \&
  \bibinfo{author}{Fert, A.}
\newblock \bibinfo{title}{Dynamics of {D}zyaloshinskii domain walls in
  ultrathin magnetic films}.
\newblock \emph{\bibinfo{journal}{EPL}} \textbf{\bibinfo{volume}{100}},
  \bibinfo{pages}{57002} (\bibinfo{year}{2012}).

\bibitem{bessarab2012harmonic}
\bibinfo{author}{Bessarab, P.~F.}, \bibinfo{author}{Uzdin, V.~M.} \&
  \bibinfo{author}{J{\'o}nsson, H.}
\newblock \bibinfo{title}{Harmonic transition-state theory of thermal spin
  transitions}.
\newblock \emph{\bibinfo{journal}{Phys. Rev. B}} \textbf{\bibinfo{volume}{85}},
  \bibinfo{pages}{184409} (\bibinfo{year}{2012}).

\bibitem{braun1994fluctuations}
\bibinfo{author}{Braun, H.-B.}
\newblock \bibinfo{title}{Fluctuations and instabilities of ferromagnetic
  domain-wall pairs in an external magnetic field}.
\newblock \emph{\bibinfo{journal}{Phys. Rev. B}} \textbf{\bibinfo{volume}{50}},
  \bibinfo{pages}{16485} (\bibinfo{year}{1994}).

\bibitem{watanabe2020counting}
\bibinfo{author}{Watanabe, H.}
\newblock \bibinfo{title}{Counting rules of {N}ambu--{G}oldstone modes}.
\newblock \emph{\bibinfo{journal}{Annu. Rev. Condens. Matter Phys.}}
  \textbf{\bibinfo{volume}{11}}, \bibinfo{pages}{169--187}
  (\bibinfo{year}{2020}).

\bibitem{kuchkin2021geometry}
\bibinfo{author}{Kuchkin, V.~M.} \emph{et~al.}
\newblock \bibinfo{title}{Geometry and symmetry in skyrmion dynamics}.
\newblock \emph{\bibinfo{journal}{Phys. Rev. B}}
  \textbf{\bibinfo{volume}{104}}, \bibinfo{pages}{165116}
  (\bibinfo{year}{2021}).

\end{thebibliography}

\section*{}
\noindent{\large{\textbf{Acknowledgement}}}\newline
We thank Hendrik Schrautzer and Tim Drevelow for valuable
discussions.
We gratefully acknowledge financial support from the Deutsche Forschungsgemeinschaft (DFG, German Research Foundation) through SPP2137 "Skyrmionics" (project no.~462602351), the
Icelandic Research Fund (Grant No.~217750), the University of Iceland Research Fund
(Grant No. 15673), and the Swedish Research Council
(Grant No. 2020-05110).

\section*{}
\noindent{\large{\textbf{Author contributions}}}\newline
M. A. G., S. M., and S. v. M. performed the MEP calculations. M. A. G., S. v. M., and P. F. B. calculated the transition rates. M. A. G. developed the treatment of the helicity and chimera rotation mode used here based on work from S. v. M., set up the master equation framework and prepared the figures. All authors contributed to the analysis and discussion of the simulation results. M. A. G., P. F. B., and S. H. wrote the first version of the manuscript and all authors contributed to the final version.

\section*{}
\noindent{\large{\textbf{Additional information}}}\newline
\noindent{\textbf{Supplementary Information}} accompanies this paper.

\noindent{\textbf{Competing interests:}} The authors declare that there are no non-financial or financial competing interests.

\end{document}